\documentclass[twocolumn,showpacs,prb]{revtex4-1}%
\usepackage{amsfonts}
\usepackage{amsmath}
\usepackage{amssymb}
\usepackage{graphicx}
\usepackage{txfonts}%
\providecommand{\U}[1]{\protect\rule{.1in}{.1in}}
\begin{document}
\title{General theory of feedback control of a nuclear spin ensemble in quantum dots}
\author{Wen Yang}
\altaffiliation{Current address: Beijing Computational Science Research Center, Beijing
100084, China}

\affiliation{Center for Advanced Nanoscience, Department of Physics, University of
California San Diego, La Jolla, California 92093-0319, USA}
\author{L. J. Sham}
\affiliation{Center for Advanced Nanoscience, Department of Physics, University of
California San Diego, La Jolla, California 92093-0319, USA}

\begin{abstract}
We present a microscopic theory of the nonequilibrium nuclear spin dynamics
driven by the electron and/or hole under continuous wave pumping in a quantum
dot. We show the correlated dynamics of the nuclear spin ensemble and the
electron and/or hole under optical excitation as a quantum feedback loop and investigate
the dynamics of the many nuclear spins as a nonlinear collective motion. This
gives rise to three observable effects: (i) hysteresis, (ii) locking
(avoidance) of the pump absorption strength to (from) the natural resonance,
and (iii) suppression (amplification) of the fluctuation of weakly polarized
nuclear spins, leading to prolonged (shortened) electron spin coherence time.
A single nonlinear feedback function as a \textquotedblleft
measurement\textquotedblright\ of the nuclear field operator in the quantum
feedback loop is constructed which determines the different outcomes of the
three effects listed above depending on the feedback being negative or
positive. The general theory also helps to put in perspective the wide range
of existing theories on the problem of a single electron spin in a nuclear
spin bath.

\end{abstract}

\maketitle

\section{Introduction}

The nonequilibrium dynamics of the nuclear spins has a long history in spin
resonance spectroscopy.\cite{AbragamBook61} The recently revived interest in
this topic is mostly due to the decoherence issue of the electron spin qubit
in semiconductor quantum dots (QDs) for quantum
computation.\cite{HansonRMP07,LaddNature10,LiuAdvPhys10} The nuclear spins,
abundant in popular III-V semiconductor QDs, produce a randomly fluctuating
nuclear field [straight arrow in Fig.~\ref{G_ENLOOP_HNLOOP}(a)] that rapidly
deprives the electron spin of its phase
coherence,\cite{KhaetskiiPRL02,MerkulovPRB02,SemenovPRB03,WitzelPRB05,WitzelPRB06,DengPRB06,YaoPRB06,YaoPRL07,LiuNJP07,YangPRB08,YangPRB08a,YangPRB09,CywinskiPRL09}
the wellspring of various advantages of quantum computation over its classical
counterpart. Suitable control of the nuclear spin dynamics can suppress the
fluctuation of the nuclear field (and hence mitigate the detrimental effect of
the electron spin decoherence) and even turn the nuclear spins into a resource
to store long-lived quantum
information.\cite{KaneNature98,TaylorPRL03,TaylorPRL03a,WitzelPRB07}

In this introduction, we introduce the most widely explored control of the
nuclear spin dynamics: dynamic nuclear polarization and more generally, the
flip of the nuclear spins by the electron and/or the hole (the removal of an
electron from the fully occupied valence band of a semiconductor). This
process, followed by the back action of the nuclear spins on the electron
and/or hole, forms different feedback loops responsible for a variety of
experimental observations, especially the suppression of the nuclear spin
fluctuation. First, in Sec.~\ref{SEC_INTRODUCTION_DNP}, we introduce dynamic
nuclear polarization, the feedback loops, and relevant experimental
observations. Then, in Sec.~\ref{SEC_INTRODUCTION_GENERALLOOP}, we briefly
survey the electron-nuclear and hole-nuclear interactions and the most general
feedback loop constructed from these interactions. Next, in
Sec.~\ref{SEC_INTRODUCTION_BACKACTION}, we summarize the exitsting theoretical
treatments of different feedback loops (especially the back action part).
Finally, in Sec.~\ref{SEC_INTRODUCTION_OURS}, we introduce our systematic,
microscopic theory of the most general feedback loop and summarize the main results.

\subsection{Dynamic nuclear polarization and feedback}

\label{SEC_INTRODUCTION_DNP}

\begin{figure}[ptb]
\includegraphics[width=\columnwidth]{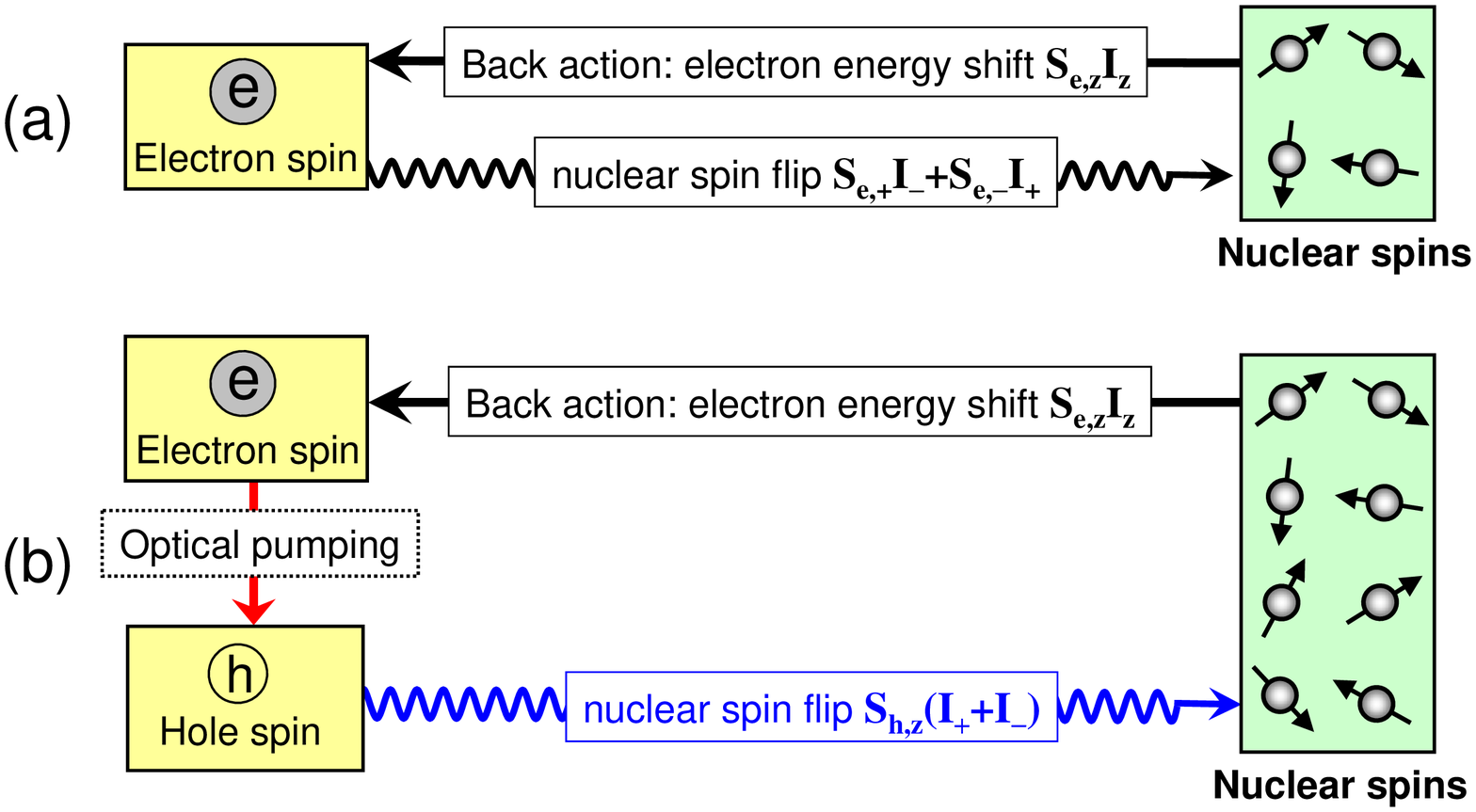} \caption{(a) Feedback
loop between the electron spin $\mathbf{\hat{S}}_{e}$ and the nuclear spins
$\{\mathbf{\hat{I}}_{j}\}$ through the contact hyperfine interaction
$\propto\mathbf{\hat{S}}_{e}\cdot\mathbf{\hat{I}}=(\hat{S}_{e,+}\hat{I}%
_{-}+\hat{S}_{e,-}\hat{I}_{+})/2+\hat{S}_{e,z}\hat{I}_{z}$: the electron flips
the nuclear spins (with or without a preferential direction) through $\hat
{S}_{e,+}\hat{I}_{-}+\hat{S}_{e,-}\hat{I}_{+}$ (wavy arrow) and changes the
nuclear field, which in turn acts back on the electron through $\hat{S}%
_{e,z}\hat{I}_{z}$ (straight arrow). (b) A specific feedback loop between the
electron spin $\mathbf{\hat{S}}_{e}$, the hole spin $\mathbf{\hat{S}}_{h}$,
and the nuclear spins $\{\mathbf{\hat{I}}_{j}\}$. First, the hole flips the
nuclear spins through the non-collinear dipolar hyperfine interaction
$\propto\hat{S}_{h,z}(\hat{I}_{+}+\hat{I}_{-})$ (wavy arrow) and changes the
nuclear field. Second, the nuclear field acts on the electron through the
diagonal part $\hat{S}_{e,z}\hat{I}_{z}$ of the contact hyperfine interaction
(straight arrow). Third, the electron is coupled to the hole through interband
optical pumping.}%
\label{G_ENLOOP_HNLOOP}%
\end{figure}

The simplest control of the nuclear spins is dynamic nuclear polarization, by
which a nonequilibrium steady-state nuclear spin polarization $s$ is induced.
Then the nuclear field acting on the electron spin [straight arrow in
Fig.~\ref{G_ENLOOP_HNLOOP}(a)] acquires a nonzero average and its fluctuation
is expected\cite{CoishPRB04} to be suppressed to $(1-s^{2})^{1/2}$ of its
thermal equilibrium fluctuation, e.g., a $\sim99\%$ nuclear spin polarization
can suppress the nuclear field fluctuation by an order of magnitude and hence
prolong the coherence time of the electron spin in the QD by the same factor.
This prospect has stimulated intensive interest in dynamic nuclear
polarization in the QD. The most widely explored scenario is to transfer the
spin angular momenta from the conduction band electron to the nuclear spins
[wavy arrow in Fig.~\ref{G_ENLOOP_HNLOOP}%
(a)]\cite{OverhauserPR53,LampelPRL1968,PagetPRB77,ImamogluPRL03,RudnerPRL07,ChristPRB07,DanonPRL08}
through the isotropic electron-nuclear contact hyperfine interaction
$\propto\mathbf{\hat{S}}_{e}\cdot\mathbf{\hat{I}}$. This scenario has been
demonstrated via different processes in various experimental setups, including
the two-electron singlet-triplet transition in transport experiments in
lateral and vertical double
QDs,\cite{OnoPRL04,KoppensScience05,DengPRB05,KoppensNature06,JouravlevPRL06,BaughPRL07,RudnerPRL07a,PettaPRL08,ReillyPRL08,FolettiNaturePhys09,GullansPRL10,ReillyPRL10,RudnerPRB10}
electron spin resonance in lateral double
QDs,\cite{LairdPRL07,NowackScience07,DanonPRL09} and in particular, interband
optical pumping in fluctuation QDs and self-assembled
QDs,\cite{BrownPRB96,GammonScience97,GammonPRL01,BrackerPRL05,YokoiPRB05,LaiPRL06,AkimovPRL06,EblePRB06,BraunPRB06,MikkelsenNaturePhys07,DzhioevPRL07,MaletinskyPRL07,TartakovskiiPRL07,OultonPRL07,MaletinskyPRB07,KajiAPL07,KajiPRB08,BelhadjPRL09,LaddPRL10,KrebsPRL10,FallahiPRL10,ChekhovichPRL10,ChekhovichPRL11}
where the highest degree of steady-state nuclear spin polarization (up to
$\sim65\%$) has been achieved.\cite{GammonPRL01,BrackerPRL05,ChekhovichPRL10}
The nonzero average nuclear field produced by the polarized nuclear spins is
then detected as an average energy shift of the electron [straight arrow in
Fig.~\ref{G_ENLOOP_HNLOOP}(a)]. In many experiments, the average energy shift
exhibits hysteretic behaviors, indicating the bistability or multistability of
the average nuclear field due to the nonlinear feedback loop
[Fig.~\ref{G_ENLOOP_HNLOOP}(a)] between the electron and the nuclear spins.
Note that, as a convention, the \textquotedblleft nuclear spin
flip\textquotedblright\ [e.g., denoted by the wavy arrow in
Fig.~\ref{G_ENLOOP_HNLOOP}(a)] may or may not have a preferential direction
and therefore is more general than dynamic nuclear polarization, which
involves nuclear spin flip with a preferential direction. 

Recently, several experimental groups reported significant suppression of the
nuclear field fluctuation for weakly or moderately polarized nuclear spins in
QD ensembles,\cite{GreilichScience07,CarterPRL09} two coupled quantum
dots,\cite{ReillyScience08,BluhmPRL2010,BarthelPRB2012,YaoEPL2010} and in
particular, single quantum
dots,\cite{XuNature09,LattaNaturePhys09,VinkNaturePhys09} an important
configuration for quantum computation. In single quantum
dots,\cite{XuNature09,LattaNaturePhys09,VinkNaturePhys09,SunPRL2012} the key
experimental observation is the maintenance (i.e., locking) of resonant
absorption over a range of pump frequency around the natural resonance. This
locking behavior arises from the shift of the electron energy level from
off-resonance to resonance by the average nuclear field. A striking
observation by both Xu \textit{et al.}\cite{XuNature09} and Latta \textit{et
al}.\cite{LattaNaturePhys09} is that the locking occurs nearly symmetrically
on both sides of the resonance. This symmetric locking reveals that the
steady-state nuclear field is antisymmetric across the resonance, a prominent
feature beyond the framework of the electron-nuclear contact hyperfine
interaction [Fig.~\ref{G_ENLOOP_HNLOOP}(a)]. In order to explain this feature,
Xu \textit{et al.}\cite{XuNature09} (followed by Ladd \textit{et
al.}\cite{LaddPRL10,LaddSPIE2011} in a different context) introduced a new
feedback loop [Fig.~\ref{G_ENLOOP_HNLOOP}(b)] consisting of the nuclear spin
flip (with no preferential direction) induced by a valence band hole inside a
trion (which consists of two inert conduction band electrons in the spin
singlet state and an unpaired valence band hole) and the back action of the
nuclear field on the conduction band electron, which is then coupled to the
hole by interband optical pumping. Very recently, the mechanism for the
hole-driven nuclear spin flip with a preferential direction (i.e., hole-driven
dynamic nuclear polarization) through the non-collinear dipolar hyperfine
interaction was also established\cite{YangPRB2012} and
generalized\cite{HogelePRL2012,UrbaszekRMP2013} to the non-collinear
electron-nuclear hyperfine interaction to explain the experimentally observed
locking and avoidance of the pump absorption strength from resonance.

\begin{figure}[ptb]
\includegraphics[width=\columnwidth]{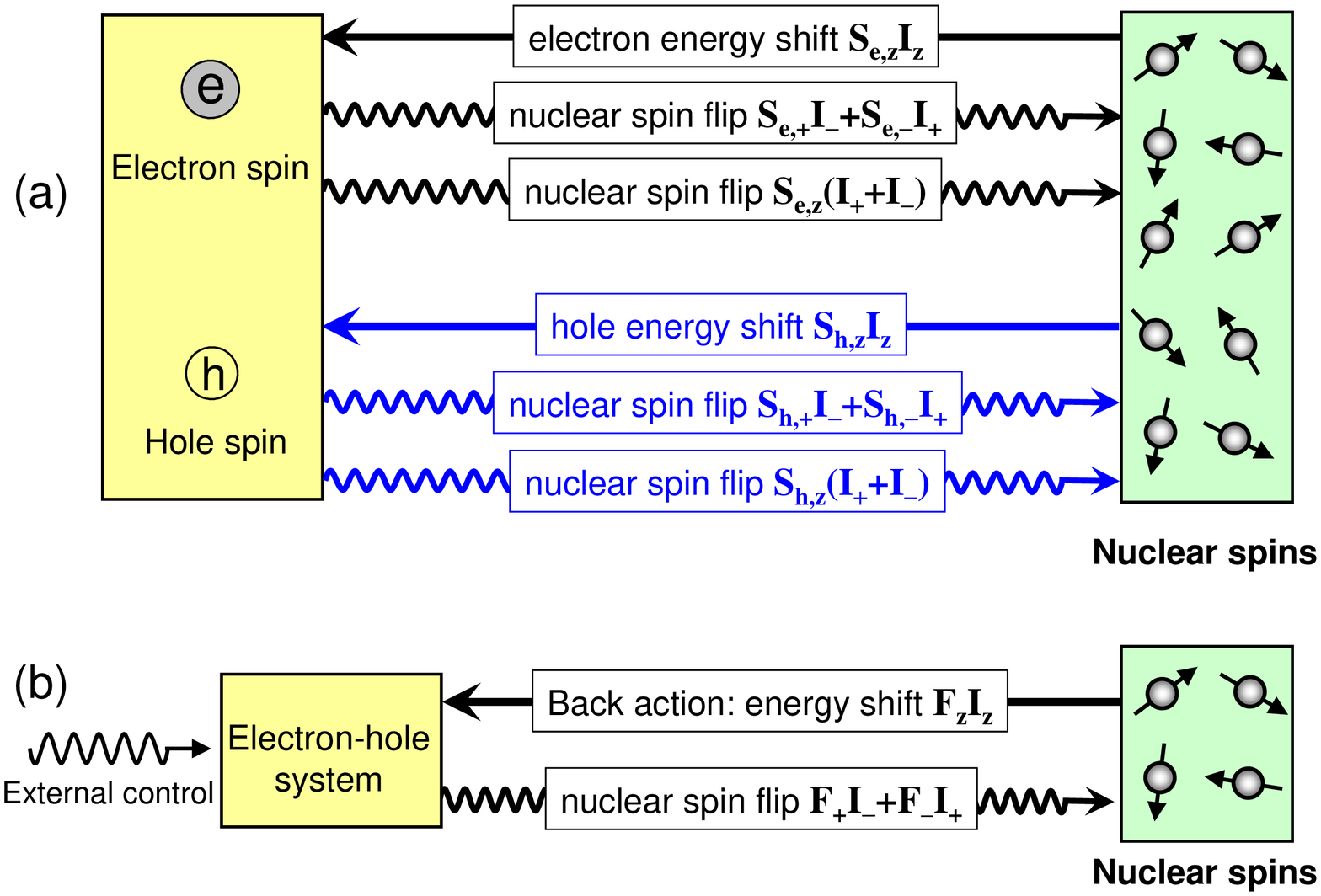} \caption{(a) Feedback
processes between the electron spin $\mathbf{\hat{S}}_{e}$, the hole spin
$\mathbf{\hat{S}}_{h}$, and the nuclear spins $\{\mathbf{\hat{I}}_{j}\}$. The
electron flips the nuclear spins through the off-diagonal contact hyperfine
interaction $\propto\hat{S}_{e,+}\hat{I}_{-}+\hat{S}_{e,-}\hat{I}_{+}$ and the
non-collinear interaction $\propto\hat{S}_{e,z}(\hat{I}_{+}+\hat{I}_{-})$. The
hole flips the nuclear spins through the off-diagonal dipolar hyperfine
interaction $\propto\hat{S}_{h,+}\hat{I}_{-}+\hat{S}_{h,-}\hat{I}_{+}$ and the
non-collinear dipolar hyperfine interaction $\propto\hat{S}_{h,z}(\hat{I}%
_{+}+\hat{I}_{-})$. The nuclear spins act back on the electron through the
diaognal contact hyperfine interaction $\propto\hat{S}_{e,z}\hat{I}_{z}$, and
on the hole through the diagonal dipolar hyperfine interaction $\propto\hat
{S}_{h,z}\hat{I}_{z}$. (b) A general feedback loop between the nuclear spins
and externally controlled electron and/or hole (hereafter referred to as e-h
system for brevity). For the sake of generality, $\hat{F}_{z}$ and $\hat
{F}_{+}$ ($=\hat{F}_{-}^{\dagger}$) refer to arbitrary electron or hole
operators (not necessarily spin operators). In particular, $\hat{F}_{z}$ does
not necessarily refer to $\hat{S}_{e,z}$ or $\hat{S}_{h,z}$ and $\hat{F}_{+}$
does not necessarily refer to $\hat{S}_{e,+}$ or $\hat{S}_{h,+}$.}%
\label{G_EHNLOOP}%
\end{figure}

\subsection{General feedback processes between electron, hole, and nuclear
spins}

\label{SEC_INTRODUCTION_GENERALLOOP}

To date, the following interactions between the electron, the hole, and the
nuclear spins have been considered:

\begin{itemize}
\item The isotropic electron-nuclear contact hyperfine interaction
$\propto\mathbf{\hat{S}}_{e}\cdot\mathbf{\hat{I}}$, which consists of the
diagonal part $\propto\hat{S}_{e,z}\hat{I}_{z}$ and the off-diagonal part
$\propto\hat{S}_{e,+}\hat{I}_{-}+\hat{S}_{e,-}\hat{I}_{+}$.

\item The anisotropic hole-nuclear dipolar hyperfine
interaction,\cite{FischerPRB08,XuNature09,EblePRL09,TestelinPRB09,FallahiPRL10,ChekhovichPRL11,YangPRB2012}
whose dominant part is diagonal $\propto\hat{S}_{h,z}\hat{I}_{z}$. Heavy-light
hole mixing\cite{BesterPRB03,KoudinovPRB04,KrizhanovskiiPRB05} introduces a
smaller off-diagonal part $\propto\hat{S}_{h,+}\hat{I}_{-}+\hat{S}_{h,-}%
\hat{I}_{+}$ and an even smaller non-collinear part $\propto\hat{S}_{h,z}%
(\hat{I}_{+}+\hat{I}_{-})$. This interaction becomes relevant when the valence
band hole is excited by interband optical pumping.

\item The non-collinear electron-nuclear hyperfine interaction $\propto\hat
{S}_{e,z}(\hat{I}_{+}+\hat{I}_{-})$, which exists between the electron of the
phosphorus donor in silicon and the $^{29}$Si isotope nuclear
spins.\cite{SaikinPRB2003,WitzelPRB07a} It may also arise in optically excited
III-V QDs when the quadrupolar axes of the nuclear spins are not parallel to
the external field.\cite{LattaPRL2011}
\end{itemize}

These interactions enable a variety of feedback processes between the
electron, the hole, and the nuclear spins [Fig.~\ref{G_EHNLOOP}(a)]. In the
general scenario, the nuclear spins can be flipped by both the electron and
the hole, through both the pair-wise flip-flop ($\hat{S}_{e,+}\hat{I}_{-}%
+\hat{S}_{e,-}\hat{I}_{+}$ or $\hat{S}_{h,+}\hat{I}_{-}+\hat{S}_{h,-}\hat
{I}_{+}$) and the non-collinear interaction [$\hat{S}_{e,z}(\hat{I}_{+}%
+\hat{I}_{-})$ or $\hat{S}_{h,z}(\hat{I}_{+}+\hat{I}_{-})$]. The nuclear spins
act back on both the electron and the hole through the diagonal interaction
($\hat{S}_{e,z}\hat{I}_{z}$ or $\hat{S}_{h,z}\hat{I}_{z}$). Such feedback
processes correlate the dynamics of different nuclear spins and play a
critical role in suppressing the nuclear field fluctuation and hence
prolonging the electron spin coherence time. In particular, all experimentally
reported
\cite{GreilichScience07,CarterPRL09,ReillyScience08,BluhmPRL2010,BarthelPRB2012,XuNature09,LattaNaturePhys09,VinkNaturePhys09}
suppressions of the nuclear field fluctuation occur in weakly or moderately
polarized systems and are attributed to feedback processes
[Fig.~\ref{G_EHNLOOP}(a)] instead of a strong nuclear spin polarization. They
demonstrate that to suppress the nuclear field fluctuation significantly,
constructing a proper feedback loop is more feasible than achieving a strong
nuclear spin polarization $s\sim99\%$, which remains an experimentally
demanding goal.

A general feedback loop [Fig.~\ref{G_EHNLOOP}(b)] between the electron and the
hole (hereafter referred to as e-h system for brevity) and the nuclear spins
consists of two steps. First, the e-h system flips the nuclear spins [wavy
arrow in Fig.~\ref{G_EHNLOOP}(b)] and hence changes the nuclear field. Second,
the nuclear field acts back on the e-h system [straight arrow in
Fig.~\ref{G_EHNLOOP}(b)]. For the first step, perturbation theory is usually
sufficient since the hyperfine interaction between the e-h system and the
nuclear spins is weak. For the second step, however, the nuclear field acting
back on the e-h system must be treated non-perturbatively since it may be
comparable with the characteristic energy scale of the electron or hole spin.

\subsection{Theoretical treatment of nuclear field back action}

\label{SEC_INTRODUCTION_BACKACTION}

In treating the nuclear field back action, many existing theories take into
account the average nuclear field but neglect its fluctuation. This approach
is capable of reproducing the average nuclear field responsible for the
experimentally observed hysteretic or locking behaviors, but provides no
information about the nuclear field fluctuation. In addition to numerical
simulation,\cite{LattaNaturePhys09,IsslerPRL10} different approaches have been
utilized to incorporate the nuclear field fluctuation:

\begin{itemize}
\item For the electron-nuclei feedback loop [Fig.~\ref{G_ENLOOP_HNLOOP}(a)],
Rudner and Levitov\cite{RudnerPRL07a,RudnerNanotechnol10} and subsequently
Danon and Nazarov\cite{DanonPRL08,DanonPRL09,VinkNaturePhys09} introduced the
stochastic approach \textit{for nuclear spin-1/2's} by assuming that the
nuclear field experiences a random walk described by a single-variable
Fokker-Planck equation. The analytical solution of the Fokker-Planck equation
quantifies the nuclear field fluctuation and shows that the competition
between dynamic nuclear polarization and nuclear spin depolarization gives
rise to a restoring force that can suppress the nuclear field fluctuation well
below the thermal equilibrium value.

\item For the electron-nuclei feedback loop [Fig.~\ref{G_ENLOOP_HNLOOP}(a)]
involving nuclear spin flip with no preferential direction, Greilich
\textit{et al.}\cite{GreilichScience07} derived a \textit{slightly different}
Fokker-Planck equation by assuming a semi-classical rate equation for the
nuclear field distribution \textit{for nuclear spin-1/2's}%
.\cite{BarnesPRL2011} The solution shows that even if the nuclear spin flip
has no preferential direction, a strong feedback suppressing the nuclear field
fluctuation can still exist in steady state, in contrast to the stochastic
approach, which gives a vanishing feedback in this case.

\item For the electron-hole-nuclei feedback loop [Fig.~\ref{G_ENLOOP_HNLOOP}%
(b)], Xu \textit{et al. }\cite{XuNature09} argued that the dependence of the
average nuclear field on the optical detuning (which in turn depends on the
fluctuating nuclear field) provides a feedback channel that can significantly
suppress the nuclear field fluctuation. This provide an intuitive, qualitative
picture for suppressing the nuclear field fluctuation by the feedback loop.

\item For the electron-hole-nuclei feedback loop [Fig.~\ref{G_ENLOOP_HNLOOP}%
(b)], our previous study\cite{YangPRB2012} established the mechanism of
hole-driven dynamic nuclear polarization through the non-collinear dipolar
hyperfine interaction [wavy arrow in Fig.~\ref{G_ENLOOP_HNLOOP}(b)]. There,
motivated by the
stochastic\cite{RudnerPRL07a,RudnerNanotechnol10,DanonPRL08,DanonPRL09,VinkNaturePhys09}
and rate equation\cite{GreilichScience07} approaches, we outlined a
microscopic derivation of the Fokker-Planck equation for this specific
mechanism without any stochastic or semi-classical assumptions. The analytical
solution quantifies the intuitive picture by Xu \textit{et al. }%
\cite{XuNature09} and establishes a connection to different
approaches.\cite{CoishPRB04,RudnerPRL07a,RudnerNanotechnol10,DanonPRL08,DanonPRL09,VinkNaturePhys09,GreilichScience07}%

\end{itemize}

The above approaches provide an excellent understanding for certain feedback
processes, but still have the drawback that they are constructed for nuclear
spin-1/2's (while the widely explored GaAs and InAs quantum dots all contain
nuclei with spins higher than 1/2) or for specific feedback loops
[Figs.~\ref{G_ENLOOP_HNLOOP}(a) and \ref{G_ENLOOP_HNLOOP}(b)] with specific
nuclear spin-flip mechanism (while the identified electron-nuclear and
hole-nuclear interactions enable more general feedback processes) and/or they
involve certain (stochastic or semi-classical) assumptions. To maximize the
control over the nuclear field and its fluctuation by flexible construction of
the feedback loop, it is desirable to develop a comprehensive understanding
for a general feedback loop and nuclear spin-flip mechanism, such as that
shown in Fig.~\ref{G_EHNLOOP}(b).

\subsection{A systematic, microscopic theory for a general feedback loop}

\label{SEC_INTRODUCTION_OURS}

In this paper, we present a systematic, microscopic theory for such a feedback
loop [Fig.~\ref{G_EHNLOOP}(b)], with the e-h system subjected to continuous
wave pumping, an important experimental situation. In particular, we study how
this feedback loop controls both the average nuclear field and its
fluctuation. This is achieved by decoupling the slow nuclear field dynamics
from the fast motion of other dynamical variables (e.g., the off-diagonal
nuclear spin coherences and the e-h variables) through the adiabatic
approximation, which enables us to incorporate non-perturbatively the back
action from the fluctuating nuclear field [straight arrow in
Fig.~\ref{G_EHNLOOP}(b)]. Our microscopic theory justifies and unifies the
stochastic
approach\cite{RudnerPRL07a,RudnerNanotechnol10,DanonPRL08,DanonPRL09,VinkNaturePhys09}
and the rate equation approach\cite{GreilichScience07} and generalizes them to
include nuclei with spins higher than 1/2. It identifies two different kinds
of steady-state feedback. The \textquotedblleft drift\textquotedblright%
\ feedback (as considered by the stochastic
approach\cite{RudnerPRL07a,RudnerNanotechnol10,DanonPRL08,DanonPRL09,VinkNaturePhys09}
and Xu \textit{et al.}\cite{XuNature09}) originates from the nonlinear drift
of the nuclear field, thus its existence requires nuclear spin flip with a
preferential direction. By contrast, the \textquotedblleft
diffusion\textquotedblright\ feedback (as considered by Greilich \textit{et
al.}\cite{GreilichScience07} and Barnes and Economou\cite{BarnesPRL2011} in
the rate equation approach and Issler \textit{et al.}%
\cite{LattaNaturePhys09,IsslerPRL10} by numerical simulation) originates from
the nonlinear diffusion of the nuclear field, so it remains efficient even
when the nuclear spin flip has no preferential direction.

In this paper we focus on the more popular \textquotedblleft
drift\textquotedblright\ feedback followed by a brief discussion about the
\textquotedblleft diffusion\textquotedblright\ feedback. The control of the
\textquotedblleft drift\textquotedblright\ feedback over the nuclear field can
be understood from three successive steps. (i) When the feedback loop is
broken by neglecting the back action, each nuclear spin is driven by the e-h
system independently. (ii) When the feedback loop is closed by taking into
account the back action from the average nuclear field, the average nuclear
field becomes coupled to the dynamics of different nuclear spins and its
motion becomes nonlinear or even multistable. This is responsible for the
experimentally observed hysteresis and absorption strength locking or
avoidance.\cite{XuNature09,LattaNaturePhys09,VinkNaturePhys09,HogelePRL2012}
(iii) When the back action from the fluctuating nuclear field is fully taken
into account, the fluctuating nuclear field becomes coupled to the dynamics of
different nuclear spins, which enables the feedback loop to further control
(e.g., suppress or amplify) the nuclear field fluctuation. This is responsible
for the experimentally observed suppression of the nuclear field fluctuation
and hence prolonged electron spin coherence
time.\cite{GreilichScience07,CarterPRL09,ReillyScience08,BluhmPRL2010,BarthelPRB2012,XuNature09,LattaNaturePhys09,VinkNaturePhys09}%

Our key finding is that all the above controls can be quantified concisely by
a single nonlinear \textit{nuclear field feedback function} $\mathbb{H}(h)$.
In the feedback loop, an \textquotedblleft input\textquotedblright\ magnetic
field $h$ from the nuclear spins [straight arrow in Fig.~\ref{G_EHNLOOP}(b)]
influences the e-h system, which in turn drives the nuclear spins [wavy arrow
in Fig.~\ref{G_EHNLOOP}(b)] to a collective mixed state, producing an
\textquotedblleft output\textquotedblright\ nuclear field $\mathbb{H}$.
Physically, this nonlinear feedback function encapsulates the mutual response
between the nuclear field and the e-h system. It provides a unified,
quantitative description to three observable effects in the steady state:

\begin{enumerate}
\item[(i)] Hysteresis, which originates from multiple stable average nuclear
fields. The average nuclear field $h^{(\mathrm{ss})}$ is determined by the
self-consistent equation $h=\mathbb{H}(h)$, which, due to the strong
nonlinearity of $\mathbb{H}(h)$, may have multiple solutions $\{h_{\alpha
}^{(\mathrm{ss})}\}$ ($\alpha=1,2,\cdots$). Each solution $h_{\alpha
}^{(\mathrm{ss})}$ is associated with a \textit{nuclear field feedback
strength}
\begin{equation}
\mathbb{H}^{\prime}(h_{\alpha}^{(\mathrm{ss})})\equiv\left(  \frac
{d\mathbb{H}(h)}{dh}\right)  _{h=h_{\alpha}^{(\mathrm{ss})}},
\label{FEEDBACK_STRENGTH}%
\end{equation}
which quantifies the sensitivity of the average \textquotedblleft
output\textquotedblright\ nuclear field to the \textquotedblleft
input\textquotedblright\ nuclear field. If $\mathbb{H}^{\prime}(h_{\alpha
}^{(\mathrm{ss})})<1$, then $h_{\alpha}^{(\mathrm{ss})}$ is a stable average
nuclear field associated with a stable feedback and a macroscopic nuclear spin state.

\item[(ii)] Locking (Avoidance) of the pump absorption strength to (from) a
certain value.\cite{XuNature09,LattaNaturePhys09,VinkNaturePhys09} Suppose
that the nuclear spins are in a macroscopic state $h_{\alpha}^{(\mathrm{ss})}$
with a feedback strength $\mathbb{H}^{\prime}(h_{\alpha}^{(\mathrm{ss})})$.
When the pump frequency $\omega$ changes by $\delta\omega$, the nuclear field
will shift the electron or hole excitation energy $\omega_{eh}$ by
$\delta\omega_{eh}$, in such a way that the detuning $\Delta\equiv\omega
_{eh}-\omega$ (which determines the pump absorption strength) changes by%
\[
\delta\Delta=\delta\omega_{eh}-\delta\omega=\frac{-\delta\omega}%
{1-\mathbb{H}^{\prime}(h_{\alpha}^{(\mathrm{ss})})}.
\]

\begin{enumerate}
\item[(ii-a)] For a strong negative feedback $\mathbb{H}^{\prime}(h_{\alpha
}^{(\mathrm{ss})})\ll-1$, we have $\left\vert \delta\Delta\right\vert
\ll\left\vert \delta\omega\right\vert $, i.e., the detuning and hence the pump
absorption strength remains nearly constant over a wide range of the pump
frequency, corresponding to the locking of the pump absorption strength to a
plateau value.

\item[(ii-b)] For a strong positive feedback $\mathbb{H}^{\prime}(h_{\alpha
}^{(\mathrm{ss})})>1$, the value $h_{\alpha}^{(\mathrm{ss})}$ becomes
unstable, leading to the avoidance of the corresponding absorption strength.

\item[(ii-c)] For a weak positive feedback $\mathbb{H}^{\prime}(h_{\alpha
}^{(\mathrm{ss})})\lesssim1$, we have $|\delta\Delta|\gg|\delta\omega|$, i.e.,
the detuning and hence the pump absorption strength changes drastically upon a
slight change of the pump frequency, corresponding to the avoidance of the
pump absorption strength from a certain value.
\end{enumerate}

\item[(iii)] The suppression or amplification of the nuclear field fluctuation
of weakly polarized nuclear spins. In a weakly polarized macroscopic nuclear
spin state $h_{\alpha}^{(\mathrm{ss})}$ with a feedback strength
$\mathbb{H}^{\prime}(h_{\alpha}^{(\mathrm{ss})})$, the feedback loop changes
the nuclear field fluctuation from the thermal equilibrium value
$\sigma_{\mathrm{eq}}$ to $\sigma_{\mathrm{eq}}[1-\mathbb{H}^{\prime
}(h_{\alpha}^{(\mathrm{ss})})]^{-1/2}$. Thus negative (positive) feedback
suppresses (amplifies) the nuclear field fluctuation. Combination of \ (ii)
and (iii) gives a positive correlation between the absorption strength locking
(avoidance) and the suppression (amplification) of the nuclear field
fluctuation: the stronger the locking (avoidance), the stronger the
suppression (amplification).
\end{enumerate}

By estimating the efficiency of the \textquotedblleft drift\textquotedblright%
\ feedback and the \textquotedblleft diffusion\textquotedblright\ feedback, we
conclude that the feedback approach is capable of suppressing the nuclear
field fluctuation to recover the intrinsic electron spin coherence time.

To exemplify our general theory, especially the quantification of the
\textquotedblleft drift\textquotedblright\ feedback by the nonlinear feedback
function, we consider the feedback loop in Fig.~\ref{G_EHNLOOP}(b), initially
proposed by Xu \textit{et al.}\cite{XuNature09} and subsequently explored by
our previous study\cite{YangPRB2012} that established the mechanism of
hole-driven dynamic nuclear polarization through the non-collinear dipolar
hyperfine interaction. This feedback loop serves as an excellent example for
our general theory because it can realize all the interesting regimes
discussed above. In particular, we find a highly nonlinear feedback function
that gives rise to bistable macroscopic nuclear spin states. For negative
nuclear Zeeman frequency, one state has a strong negative feedback
$\mathbb{H}^{\prime}(h_{\alpha}^{(\mathrm{ss})})\ll-1$, leading to strong
locking of the pump absorption strength to the resonance and significantly
suppressed nuclear field fluctuation. When the nuclear Zeeman frequency is
reversed, one state has a positive feedback$,$ leading to strong avoidance of
the pump absorption strength from resonance and enhanced nuclear field fluctuation.

\section{Theory}

\label{SEC_THEORY}

We consider many nuclear spins coupled to a generic e-h system under
continuous wave pumping in a single QD subjected to an external magnetic field
$B$ along the $z$ growth axis. The total Hamiltonian is%
\begin{equation}
\hat{H}(t)=\hat{H}_{N}+\hat{H}_{eh}(t)+\hat{V}(t). \label{HAMILTONIAN}%
\end{equation}
The nuclear spin Hamiltonian is
\begin{equation}
\hat{H}_{N}\equiv\sum_{j}\omega_{j,N}\hat{I}_{j}^{z}, \label{HN}%
\end{equation}
where $\omega_{j,N}\equiv-\gamma_{j,N}B$ is the nuclear Zeeman frequency and
the summation $\sum_{j}$ runs over all nuclear spins in the QD. The e-h
Hamiltonian $\hat{H}_{eh}(t)$ includes the continuous pumping and the coupling
of the e-h system to the environment (e.g., vacuum electromagnetic
fluctuation\cite{ScullyQuantumOptics} or neighboring electron/hole
reservoirs\cite{DreiserPRB08}), which introduces damping into the e-h system.
The general coupling between the e-h system and the nuclear spins can be
written as%
\begin{equation}
\hat{V}(t)\equiv\hat{F}_{z}(t)\hat{h}_{z}+\hat{F}_{+}(t)\hat{h}_{-}+\hat
{F}_{-}(t)\hat{h}_{+}, \label{V}%
\end{equation}
where $\hat{F}_{z}(t)=\hat{F}_{z}^{\dagger}(t)$ and $\hat{F}_{-}(t)=\hat
{F}_{+}^{\dagger}(t)$ are arbitrary dimensionless operators (not necessarily
spin operators) for the electron or the hole. In particular, $\hat{F}_{z}$
does not necessarily refers to $\hat{S}_{e,z}$ or $\hat{S}_{h,z}$ and $\hat
{F}_{+}$ does not necessarily refer to $\hat{S}_{e,+}$ or $\hat{S}_{h,+}$.
These operators are coupled to different components
\begin{align*}
\hat{h}_{z}  &  \equiv\sum_{j}a_{j,z}\hat{I}_{j,z},\\
\hat{h}_{+}  &  \equiv\sum_{j}a_{j,+}\hat{I}_{j,+},
\end{align*}
and $\hat{h}_{-}=\hat{h}_{+}^{\dagger}$ of the nuclear field, where $\hat
{I}_{j,\pm}\equiv\hat{I}_{j,x}\pm i\hat{I}_{j,y}$. The feedback loop in this
model corresponds to Fig.~\ref{G_EHNLOOP}(b) with $\hat{I}_{\alpha}$ replaced
by $\hat{h}_{\alpha}$. Through the off-diagonal coupling
\begin{equation}
\hat{V}_{\mathrm{nd}}(t)\equiv\hat{F}_{+}(t)\hat{h}_{-}+\hat{F}_{-}(t)\hat
{h}_{+}, \label{VND}%
\end{equation}
the e-h system flips the nuclear spins [wave arrow in Fig.~\ref{G_EHNLOOP}(b)]
and changes the nuclear field, which in turn acts back on the e-h system
through the diagonal coupling $\hat{F}_{z}(t)\hat{h}_{z}$ [straight arrow in
Fig.~\ref{G_EHNLOOP}(b)]. To incorporate non-perturbatively the back action by
the diagonal coupling, we divide the total Hamiltonian $\hat{H}(t)$ into the
diagonal, unperturbed part
\begin{equation}
\hat{H}_{0}(t)\equiv\hat{H}_{N}+\hat{H}_{eh}(t)+\hat{F}_{z}(t)\hat{h}_{z},
\label{H0}%
\end{equation}
to be treated non-perturbatively, and the off-diagonal part $\hat
{V}_{\mathrm{nd}}(t)$, to be treated perturbatively.

We are interested in the control of the feedback loop over the nuclear field
dynamics, which is associated with the diagonal part $\hat{P}(t)$ of the
nuclear spin density matrix. Therefore, we need to single out the motion of
$\hat{P}(t)$ from the exact equation of motion
\[
\frac{d}{dt}\hat{\rho}(t)=-i[\hat{H}_{0}(t)+\hat{V}_{\mathrm{nd}}(t),\hat
{\rho}(t)]
\]
for the density matrix $\hat{\rho}(t)$ of the coupled system. This can be
achieved by the following time-scale analysis for three essential processes,
two being driven by the unperturbed Hamiltonian $\hat{H}_{0}(t)$ and one being
driven by the perturbation $\hat{V}_{\mathrm{nd}}(t)$:

\begin{enumerate}
\item Dissipative dynamics of the e-h system driven by $\hat{H}_{0}(t)$. Here
$\hat{h}_{z}$ may be regarded as a classical parameter since it commutes with
every term in $\hat{H}_{0}(t)$. Through the diagonal coupling $\hat{F}%
_{z}(t)\hat{h}_{z}$ in $\hat{H}_{0}(t)$, the back action of the nuclear field
$\hat{h}_{z}$ on the e-h system [straight arrow in Fig.~\ref{G_EHNLOOP}(b)]
changes the free e-h evolution
\begin{equation}
\hat{U}_{eh}(t)=\mathcal{T}e^{-i\int_{0}^{t}\hat{H}_{eh}(t^{\prime}%
)dt^{\prime}} \label{UEH}%
\end{equation}
to a $\hat{h}_{z}$-dependent evolution
\begin{equation}
\hat{U}_{eh}(\hat{h}_{z},t)=\mathcal{T}e^{-i\int_{0}^{t}[\hat{H}%
_{eh}(t^{\prime})+\hat{F}_{z}(t^{\prime})\hat{h}_{z}]dt^{\prime}}
\label{UEH_EFF}%
\end{equation}
($\mathcal{T}$ is the time-ordering operator) that establishes a $\hat{h}_{z}%
$-dependent steady e-h state $\hat{\rho}_{eh}^{(\mathrm{ss})}(\hat{h}%
_{z},t)=\hat{U}_{eh}(\hat{h}_{z},t)\hat{\rho}_{eh}(0)\hat{U}_{eh}^{\dagger
}(\hat{h}_{z},t)$ within the e-h relaxation time $T_{eh}\sim1\ \mathrm{ns}%
$\cite{XuNature09} [recall that $\hat{H}_{eh}(t)$ includes the e-h relaxation].

\item Nuclear spin dephasing driven by $\hat{H}_{0}(t)$. Through the diagonal
coupling in $\hat{H}_{0}(t)$, the e-h fluctuation eliminates the off-diagonal
nuclear spin coherences (see Appendix \ref{APPENDIX_DEPHASING} for details)
within the nuclear spin dephasing time $T_{2,N}\sim0.01-1\ \mathrm{ms}$.
Additional nuclear spin dephasing on the time scale $\sim0.1\ \mathrm{ms}$
comes from the nuclear-nuclear dipolar
interaction.\cite{AbragamBook61,WitzelPRB07} This process transforms an
arbitrary nuclear spin density matrix $\hat{\rho}_{N}(t)$ to a diagonal one
$\hat{P}(t)$ with vanishing nuclear spin coherences.

\item Nuclear spin relaxation driven by $\hat{V}_{\mathrm{nd}}(t)$. Through
the off-diagonal coupling, the e-h fluctuation flips the nuclear spins [wavy
arrow in Fig.~\ref{G_EHNLOOP}(b)] and changes the nuclear field within the
nuclear spin relaxation time $T_{1,N}\sim1-100$ $\mathrm{s}$%
.\cite{OnoPRL04,LaiPRL06,LairdPRL07,MikkelsenNaturePhys07,GreilichScience07,ChekhovichPRL10}
The decay of the nuclear field due to, e.g., the non-secular part of the
nuclear-nuclear dipolar interaction, occurs on the same time
scale.\cite{MaletinskyPRL07,BaughPRL07,GreilichScience07,ReillyScience08}
\end{enumerate}

\begin{figure}[ptb]
\includegraphics[width=\columnwidth]{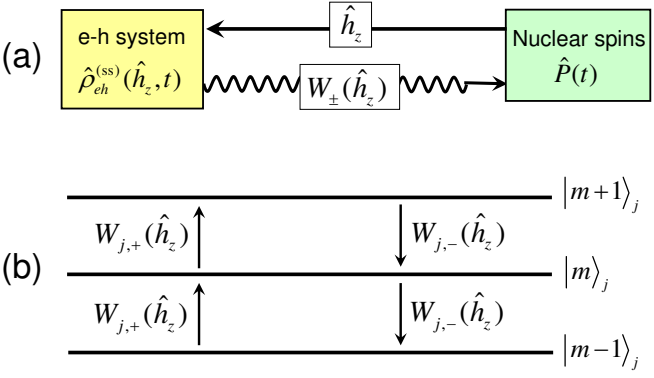}\caption{(a)\ The back
action of the nuclear field $\hat{h}_{z}$ (straight arrow) instantaneously
establish a $\hat{h}_{z}$-dependent e-h steady state $\hat{\rho}%
_{eh}^{(\mathrm{ss})}(\hat{h}_{z},t)$, whose fluctuation in turn flips the
nuclear spin (wavy arrow) and changes the nuclear field. (b) The spin-flip
transition rates $W_{j,\pm}(\hat{h}_{z})$ of the $j$th nuclear spin between
adjacent eigenstates $\{\left\vert m\right\rangle _{j}\}$ of $\hat{I}_{j}^{z}$
are dependent on the nuclear field $\hat{h}_{z}$.}%
\label{G_SingleSpinJump}%
\end{figure}

To summarize, the unperturbed evolution driven by $\hat{H}_{0}(t)$ rapidly
establish a classically correlated state $\hat{\rho}_{eh}^{(\mathrm{ss})}%
(\hat{h}_{z},t)\hat{P}(t)$ on a short time scale $\sim T_{eh},T_{2,N}$, while
the off-diagonal coupling $\hat{V}_{\mathrm{nd}}(t)$ slowly flips the nuclear
spins and changes the nuclear field $\hat{h}_{z}$ on a much longer time scale
$\sim T_{1,N}\gg T_{eh},T_{2,N}$. This fact enables us to use the adiabatic
approximation to separate the unperturbed evolution from the nuclear spin flip
by assuming that the classically correlated state is instantaneously
established after each nuclear spin flip. In this case, on the time scale of
$T_{1,N}$, we can use the classically correlated state $\hat{\rho}%
_{eh}^{(\mathrm{ss})}(\hat{h}_{z},t)\hat{P}(t)$ as the zeroth-order
approximation to the state of the whole system and incorporate the
off-diagonal coupling [wavy arrow in Fig.~\ref{G_EHNLOOP}(b)] through the
second-order perturbation theory in the density matrix formalism. Through
straightforward algebra (see Appendix \ref{APPENDIX_RELAXATION} for details),
we arrive at the following rate equation for $\hat{P}(t)$ up to second order
of $\hat{V}_{\mathrm{nd}}$ on the time scale of $T_{1,N}$ under the condition
that the e-h fluctuation is invariant under temporal translation:%
\begin{equation}
\frac{d}{dt}\hat{P}(t)=-\sum_{j}[\hat{I}_{j,-},\hat{I}_{j,+}W_{j,+}(\hat
{h}_{z})\hat{P}(t)]-\sum_{j}[\hat{I}_{j,+},\hat{I}_{j,-}W_{j,-}(\hat{h}%
_{z})\hat{P}(t)], \label{RATEEQ}%
\end{equation}
where
\begin{equation}
W_{j,\pm}(\hat{h}_{z})=\left\vert a_{j,+}\right\vert ^{2}\int_{-\infty
}^{\infty}e^{\mp i\omega_{j,N}t}dt\ \operatorname*{Tr}\nolimits_{eh}\hat
{F}_{\pm}^{\mathrm{I}}(\hat{h}_{z},t)\hat{F}_{\mp}^{\mathrm{I}}(\hat{h}%
_{z},0)\hat{\rho}_{eh}^{(\mathrm{ss})}(\hat{h}_{z},0) \label{WPN}%
\end{equation}
is the rate of the transition of the $j$th nuclear spin that increases [for
$W_{j,+}(\hat{h}_{z})$] or decreases [for $W_{j,-}(\hat{h}_{z})$] the quantum
number of $\hat{I}_{j,z}$ by one [Fig.~\ref{G_SingleSpinJump}(b)] and
\begin{equation}
\hat{O}^{\mathrm{I}}(\hat{h}_{z},t)\equiv\hat{U}_{eh}(\hat{h}_{z},t)\hat
{O}(t)\hat{U}_{eh}^{\dagger}(\hat{h}_{z},t) \label{O_INT}%
\end{equation}
for an arbitrary e-h operator $\hat{O}(t)$. The e-h fluctuation
$\operatorname*{Tr}\nolimits_{eh}\hat{F}_{\pm}^{\mathrm{I}}(\hat{h}_{z}%
,t)\hat{F}_{\mp}^{\mathrm{I}}(\hat{h}_{z},0)\hat{\rho}_{eh}^{(\mathrm{ss}%
)}(\hat{h}_{z},0)$ and hence the transition rates $W_{j,\pm}(\hat{h}_{z})$ can
be evaluated through the quantum regression theorem.\cite{ScullyQuantumOptics}

Equation (\ref{RATEEQ}) describes the dynamics of the diagonal part of the
nuclear spin density matrix (i.e., the population flow of the nuclear spins)
driven by the feedback loop. Equation (\ref{WPN}) is the non-equilibrium
version\cite{DanonPRB2011} of the fluctuation-dissipation
theorem:\cite{GardinerQuantumNoise04} the fluctuation of the non-equilibrium
e-h system [driven by the $\hat{h}_{z}$-dependent evolution $\hat{U}_{eh}%
(\hat{h}_{z},t)$] induces irreversible population flow of the nuclear spins
towards a non-equilibrium steady state. Now the entire feedback loop reduces
from Fig.~\ref{G_EHNLOOP}(b) to Fig.~\ref{G_SingleSpinJump}(a). First, the
\textquotedblleft input\textquotedblright\ nuclear field $\hat{h}_{z}$ acting
on the e-h system [straight arrow in Fig.~\ref{G_SingleSpinJump}(a)] changes
the free e-h evolution $\hat{U}_{eh}(t)$ [Eq.~(\ref{UEH})] to a $\hat{h}_{z}%
$-dependent evolution $\hat{U}_{eh}(\hat{h}_{z},t)$ [Eq.~(\ref{UEH_EFF})] that
instantaneously establishes the $\hat{h}_{z}$-dependent e-h steady state
$\hat{\rho}_{eh}^{(\mathrm{ss})}(\hat{h}_{z},t)$ and hence e-h fluctuation
$\operatorname*{Tr}\nolimits_{eh}\hat{F}_{\pm}^{\mathrm{I}}(\hat{h}_{z}%
,t)\hat{F}_{\mp}^{\mathrm{I}}(\hat{h}_{z},0)\hat{\rho}_{eh}^{(\mathrm{ss}%
)}(\hat{h}_{z},0)$. Second, through the off-diagonal coupling [wavy arrow in
Fig.~\ref{G_SingleSpinJump}(a)], the $\hat{h}_{z}$-dependent e-h fluctuation
induces an $\hat{h}_{z}$-dependent irreversible population flow of the nuclear
spins [Fig.~\ref{G_SingleSpinJump}(b)]. Then the \textquotedblleft
output\textquotedblright\ nuclear field generated by this population flow
depends on the \textquotedblleft input\textquotedblright\ nuclear field
$\hat{h}_{z}$. Finally, the back action of this \textquotedblleft
output\textquotedblright\ nuclear field on the e-h system [straight arrow in
Fig.~\ref{G_SingleSpinJump}(a)] closes the feedback loop.

Up to now we have neglected the nuclear spin depolarization, e.g., by the
nuclear-nuclear dipolar interactions. If these processes do not interfere with
the e-h mechanism considered here, then they can be characterized by
phenomenological decay rates $\{\Gamma_{j,1}\}$ and incorporated into
Eq.~(\ref{RATEEQ}) by replacing the transition rates $W_{j,\pm}(\hat{h}_{z})$
by $W_{j,\pm}(\hat{h}_{z})+\Gamma_{j,1}/2$. Hereafter it is understood that
$W_{j,\pm}(\hat{h}_{z})$ already includes the nuclear spin depolarization.

Our solution of Eq. (\ref{RATEEQ}) consists in the control of the average
nuclear field and the nuclear field fluctuation by the feedback loop. For
simplicity we consider uniform couplings $a_{j,+}=a_{+},a_{j,z}=a_{z}$
(generalization to non-uniform couplings can be achieved by coarse
graining\cite{PetrovPRB09}) of the e-h system to identical nuclear spin-$I$'s
with $\omega_{j,N}=\omega_{N}$ in the QD, so that $W_{j,\pm}(\hat{h}%
_{z})=W_{\pm}(\hat{h}_{z})$ is independent of $j$. The nuclear field
\[
\hat{h}_{z}=\left(  NIa_{z}\right)  \times\left(  \frac{1}{NI}\sum_{j=1}%
^{N}\hat{I}_{j}^{z}\right)  \equiv h_{\max}\hat{s},
\]
where $N$ is the total number of nuclear spins in the QD, $h_{\max}\equiv
NIa_{z}$ is the nuclear field from fully polarized nuclear spins, and $\hat
{s}\equiv\hat{h}_{z}/h_{\max}$ is the polarization per unit nuclear spin or
the normalized nuclear field. Hereafter, we will refer to $\hat{s}$ as the
nuclear field in cases of no confusion.

The explanation of the feedback control is organized as follows. (A) We start
with a nuclear field operator $\hat{h}_{z}$ acting back on the e-h system
[straight arrow in Fig.~\ref{G_SingleSpinJump}(a)] taking a constant value $h$
(as if it were measured) and introduce the notion of a \textit{nuclear field
feedback function} $\mathbb{H}(h)$. (B) We take into account the back action
of the average nuclear field, discuss the multistability of the nuclear field,
and use the \textit{nuclear field feedback strength} $\mathbb{H}^{\prime
}(h)\equiv d\mathbb{H}(h)/dh$ to quantify the locking (or avoidance) of the
pump absorption strength to (or from) a certain value. (C) We close the
feedback loop by fully taking into account the back action from the
fluctuating nuclear field $\hat{h}_{z}$. In this case, we identify two
different kinds of steady-state feedback: the \textquotedblleft
drift\textquotedblright\ feedback originating from the nonlinear drift of the
nuclear field and the \textquotedblleft diffusion\textquotedblright\ feedback
originating from the nonlinear diffusion of the nuclear field. We show that
the nuclear field fluctuation controlled by the \textquotedblleft
drift\textquotedblright\ feedback is quantified by the nuclear spin
polarization\cite{CoishPRB04} (negligible for weakly polarized system) and the
feedback strength $\mathbb{H}^{\prime}(h)$. By estimating the efficiency of
the \textquotedblleft drift\textquotedblright\ feedback and \textquotedblleft
diffusion\textquotedblright\ feedback, we conclude that the feedback approach
is capable of suppressing the nuclear field fluctuation to recover the
intrinsic electron spin coherence time.

\subsection{Back action from constant nuclear field: nuclear field feedback
function}

\label{SEC_CONSTANT_FEEDBACK}

Here we assume that the nuclear field $\hat{h}_{z}$ acting on the e-h system
takes a constant value $h$, then the e-h induced nuclear spin-flip rate
$W_{\pm}(\hat{h}_{z})\rightarrow W_{\pm}(h)$ becomes $c$-numbers as the
feedback loop is \textquotedblleft measured\textquotedblright. In this case,
the dynamics of different nuclear spins are decoupled and the density matrix
for all the nuclear spins is the product of the density matrices of individual
nuclear spins. The average polarization $\langle\hat{I}_{j}^{z}\rangle/I$ of
each nuclear spin is equal to $s(t)\equiv\operatorname*{Tr}\hat{s}\hat{P}(t)$.
Therefore, as long as $s(t)$ is concerned, we need only consider one nuclear
spin in this case.

For nuclear spin-1/2's, Eq.~(\ref{RATEEQ}) gives%
\begin{equation}
\frac{d}{dt}s=-\Gamma_{\mathrm{tot}}(h)[s-s_{0}^{(1/2)}(h)], \label{EOM_S12}%
\end{equation}
where%
\begin{equation}
s_{0}^{(1/2)}(h)=\frac{W_{+}(h)-W_{-}(h)}{W_{+}(h)+W_{-}(h)} \label{S0}%
\end{equation}
is the steady-state nuclear spin polarization, established within the nuclear
spin relaxation time $T_{1,N}(h)=1/\Gamma_{\mathrm{tot}}(h)$, where
$\Gamma_{\mathrm{tot}}(h)\equiv W_{+}(h)+W_{-}(h)$.

For nuclei with a general spin $I$, the steady-state nuclear spin polarization
becomes%
\begin{equation}
s_{0}(h)\equiv B_{I}\left(  I\ln\frac{1+s_{0}^{(1/2)}(h)}{1-s_{0}^{(1/2)}%
(h)}\right)  \approx\frac{2(I+1)}{3}s_{0}^{(1/2)}(h), \label{S0I}%
\end{equation}
where $B_{I}(x)\equiv(2I+1)/(2I)\coth[(2I+1)x/(2I)]-1/(2I)\coth[x/(2I)]$ is
the Brillouin function. The real-time motion of $s(t)$ is given by%
\begin{align}
\frac{d}{dt}s  &  =-\Gamma_{\mathrm{tot}}(h)\left(  s-\frac{\langle\hat
{I}_{j,x}^{2}\rangle+\langle\hat{I}_{j,y}^{2}\rangle}{I}s_{0}^{(1/2)}%
(h)\right) \nonumber\\
&  \approx-\Gamma_{\mathrm{tot}}(h)[s-s_{0}(h)]. \label{EOM_SI}%
\end{align}
The last step of Eqs.~(\ref{S0I}) and (\ref{EOM_SI}) is valid if $I=1/2$ or
$|s_{0}(h)|\ll1$. \textit{Below, unless explicitly stated, we always consider
this situation.} Equation (\ref{S0I}) shows that for weak polarization
$|s_{0}(h)|\ll1$, the polarization of nuclear spin-$I$'s is enhanced by a
factor $\sim2(I+1)/3$ compared with that of nuclear spin-1/2's.

The steady state value of the average nuclear field $\operatorname*{Tr}\hat
{h}_{z}\hat{P}(t)$ as a function of $h$ is given by a nonlinear function
\begin{equation}
\mathbb{H}(h)\equiv h_{\max}s_{0}(h). \label{FEEDBACK_FUNCTION}%
\end{equation}
Since the function $\mathbb{H}(h)$ connects the \textquotedblleft
input\textquotedblright\ nuclear field $h$ acting on the e-h system [straight
arrow in Fig.~\ref{G_SingleSpinJump}(a)] and the average \textquotedblleft
output\textquotedblright\ nuclear field produced by the nuclear spins driven
by the e-h fluctuation [wavy arrow in Fig.~\ref{G_SingleSpinJump}(a)], we call
$\mathbb{H}(x)$ the\textit{ nuclear field feedback function}. It encapsulates
(i) the nonlinear response of the e-h fluctuation to the nuclear field acting
on the e-h system [straight arrow in Fig.~\ref{G_SingleSpinJump}(a)] and
(ii)\ the response of the nuclear field to the e-h fluctuation [wavy arrow in
Fig.~\ref{G_SingleSpinJump}(b)]. Equation (\ref{FEEDBACK_FUNCTION}) also shows
that $s_{0}(h)=\mathbb{H}(h)/h_{\max}$ is just the normalized nuclear field
feedback function.

\subsection{Back action from average nuclear field: absorption strength
locking or avoidance}

\label{SEC_MEANFIELD_FEEDBACK}

Here we take into account the average nuclear field $h(t)\equiv
\operatorname*{Tr}\hat{P}(t)\hat{h}_{z}$ acting on the e-h system, i.e.,
$W_{\pm}(\hat{h}_{z})\rightarrow W_{\pm}(h(t))$. In this case, the dynamics of
different nuclear spins are coupled to the average nuclear field $h(t)$. This
enables the feedback loop to control the average nuclear field. As a result,
the motion of $h(t)=h_{\max}s(t)$, as obtained from Eq.~(\ref{EOM_SI}) by
replacing $h$ with $h(t)$, becomes nonlinear:%

\begin{equation}
\frac{d}{dt}h(t)\approx-\Gamma_{\mathrm{tot}}(h(t))[h(t)-\mathbb{H}(h(t))].
\label{EOM_MEANFIELD}%
\end{equation}

The average nuclear field $h^{(\mathrm{ss})}$ in the steady state is
determined by the self-consistent equation $h=\mathbb{H}(h)$, which, due to
the nonlinearity of $\mathbb{H}(h)$, may have multiple solutions $\{h_{\alpha
}^{(\mathrm{ss})}\}$ (distinguished by the subscript $\alpha=1,2,\cdots$).
Each solution $h_{\alpha}^{(\mathrm{ss})}$ is associated with a
\textit{nuclear field feedback strength} $\mathbb{H}^{\prime}(h_{\alpha
}^{(\mathrm{ss})})$, as defined in Eq.~(\ref{FEEDBACK_STRENGTH}). A positive
(negative) feedback corresponds to $\mathbb{H}^{\prime}(h_{\alpha
}^{(\mathrm{ss})})>0$ [$\mathbb{H}^{\prime}(h_{\alpha}^{(\mathrm{ss})})<0$].
The equation of motion for the deviation $\delta h_{\alpha}(t)\equiv
h(t)-h_{\alpha}^{(\mathrm{ss})}$ of the average nuclear field $h(t)$ from the
$\alpha$-th steady-state value $h_{\alpha}^{(\mathrm{ss})}$ follows from
Eq.~(\ref{EOM_MEANFIELD}) as%
\begin{equation}
\frac{d}{dt}\delta h_{\alpha}\approx-\Gamma_{\mathrm{tot}}(h_{\alpha
}^{(\mathrm{ss})})[1-\mathbb{H}^{\prime}(h_{\alpha}^{(\mathrm{ss})})]\delta
h_{\alpha}+O[(\delta h_{\alpha})^{2}]. \label{FLUCTUATION}%
\end{equation}
For $h_{\alpha}^{(\mathrm{ss})}$ to be stable, the corresponding feedback
strength must satisfy $\mathbb{H}^{\prime}(h_{\alpha}^{(\mathrm{ss})})<1$, so
that any deviation of the average nuclear field away from its steady-state
value $h_{\alpha}^{(\mathrm{ss})}$ would decay to zero within the nuclear spin
relaxation time $T_{1,N}(h_{\alpha}^{(\mathrm{ss})})=1/\Gamma_{\mathrm{tot}%
}(h_{\alpha}^{(\mathrm{ss})})$. In this case, $h_{\alpha}^{(\mathrm{ss})}$
corresponds to a macroscopic nuclear spin state. For weak nuclear spin
polarization, since $s_{0}\approx2(I+1)s_{0}^{(1/2)}/3$ and $\mathbb{H}%
=h_{\max}s_{0}=Na_{z}Is_{0}$, we have $\mathbb{H}^{\prime}(h_{\alpha
}^{(\mathrm{ss})})\propto I(I+1)$, i.e., nuclei with higher spin have stronger
feedback strength.

Recently, under continuous wave pumping, several
groups\cite{XuNature09,LattaNaturePhys09,VinkNaturePhys09} observed the
locking of the pump absorption strength to the resonance: when gradually
sweeping the pump frequency $\omega$ away from the resonance with the electron
or hole excitation, the nuclear field tends to compensate this change and
shift the electron or hole excitation energy to restore the resonance. Very
recently, the opposite behavior (i.e., pushing the pump absorption strength
away from the natural resonance) was predicted\cite{YangPRB2012} and
observed.\cite{HogelePRL2012} These behaviors originate from the feedback of
the average nuclear field. Below we use the nuclear field feedback function to
quantify these (and more general) behaviors.

In a typical continuous pumping experiment, the back action of an average
nuclear field $h$ on the e-h system shifts the electron or hole excitation
energy from $\omega_{eh}^{0}$ to $\omega_{eh}\equiv\omega_{eh}^{0}+h$ or
$\omega_{eh}\equiv\omega_{eh}^{0}-h$. For specificity we first consider the
former case $\omega_{eh}=\omega_{eh}^{0}+h$. The detuning between the electron
or hole excitation energy and the pump frequency is $\Delta\equiv\omega
_{eh}-\omega=\omega_{eh}^{0}+h-\omega$. Typically the nuclear spin transition
rates $W_{\pm}$ (due to e-h fluctuation) are nonlinear functions of $\Delta$
and this is the only source for the dependence of the nuclear field feedback
function on $h$. In this case, the feedback function can be written as
$\mathbb{H}(h)=\mathcal{H}(\omega_{eh}^{0}+h-\omega)$. Suppose that at an
initial pump frequency $\omega$, the nuclear spins are in the $\alpha$-th
macroscopic state with an average nuclear field $h_{\alpha}^{(\mathrm{ss})}$
determined by%
\begin{equation}
h_{\alpha}^{(\mathrm{ss})}=\mathcal{H}(\omega_{eh}^{0}+h_{\alpha
}^{(\mathrm{ss})}-\omega), \label{tmp1}%
\end{equation}
the electron or hole excitation energy is $\omega_{eh}=\omega_{eh}%
^{0}+h_{\alpha}^{(\mathrm{ss})}$, and the detuning is $\Delta=\omega_{eh}%
^{0}+h_{\alpha}^{(\mathrm{ss})}-\omega$. Now the pump frequency changes by
$\delta\omega$ (which is not necessarily small), then the nuclear field
changes by $\delta h_{\alpha}^{(\mathrm{ss})}$ determined by%
\begin{equation}
h_{\alpha}^{(\mathrm{ss})}+\delta h_{\alpha}^{(\mathrm{ss})}=\mathcal{H}%
(\omega_{eh}^{0}+h_{\alpha}^{(\mathrm{ss})}+\delta h_{\alpha}^{(\mathrm{ss}%
)}-\omega-\delta\omega), \label{tmp2}%
\end{equation}
the electron or hole excitation energy changes by $\delta\omega_{eh}=\delta
h_{\alpha}^{(\mathrm{ss})}$, and the detuning changes by $\delta\Delta
=\delta\omega_{eh}-\delta\omega$. If the detuning change $\delta\Delta$ is
small, then we can make a first-order Taylor expansion to $\mathcal{H}%
(\omega_{eh}^{0}+h_{\alpha}^{(\mathrm{ss})}+\delta h_{\alpha}^{(\mathrm{ss}%
)}-\omega-\delta\omega)=\mathcal{H}(\omega_{eh}^{0}+h_{\alpha}^{(\mathrm{ss}%
)}-\omega+\delta\Delta)$ and obtain%
\begin{align}
\delta\omega_{eh}  &  =-\frac{\mathbb{H}^{\prime}(h_{\alpha}^{(\mathrm{ss})}%
)}{1-\mathbb{H}^{\prime}(h_{\alpha}^{(\mathrm{ss})})}\delta\omega
,\label{INC_WEH}\\
\delta\Delta &  =\frac{-\delta\omega}{1-\mathbb{H}^{\prime}(h_{\alpha
}^{(\mathrm{ss})})}. \label{INC_DEFF}%
\end{align}
For the nuclear field shifting the electron or hole excitation energy from
$\omega_{eh}^{0}$ to $\omega_{eh}\equiv\omega_{eh}^{0}-h$, Eqs.~(\ref{INC_WEH}%
) and (\ref{INC_DEFF}) still hold.

Equation (\ref{INC_DEFF}) shows that the feedback loop controls the
sensitivity of the pump detuning $\Delta$ [and hence the pump absorption
strength $\chi(\Delta)$] to the change of the pump frequency (for clarity we
assume that the nuclear field shifts the electron or hole excitation energy
from $\omega_{eh}^{0}$ to $\omega_{eh}\equiv\omega_{eh}^{0}+h$ although the
same conclusions apply to the opposite case $\omega_{eh}\equiv\omega_{eh}%
^{0}-h$):

\begin{itemize}
\item If $h_{\alpha}^{(\mathrm{ss})}$ is associated with a strong negative
feedback $\mathbb{H}^{\prime}(h_{\alpha}^{(\mathrm{ss})})\ll-1$, then the
nuclear field induced shift of the electron or hole excitation energy
$\delta\omega_{eh}\approx\delta\omega$ [Eq.~(\ref{INC_WEH})] largely
compensates the change of the pump frequency, so that the change of the
detuning $\left\vert \delta\Delta\right\vert =\left\vert \delta\omega
_{eh}-\delta\omega\right\vert \ll\left\vert \delta\omega\right\vert $
[Eq.~(\ref{INC_DEFF})] is very small, which in turn justifies the first-order
Taylor expansion to $\mathcal{H}(\omega_{eh}^{0}+h_{\alpha}^{(\mathrm{ss}%
)}-\omega+\delta\Delta)$ [used to derive Eqs.~(\ref{INC_WEH}) and
(\ref{INC_DEFF})] even when $\delta\omega$ is not small. As a result, the
absorption strength becomes insensitive to the change of the pump frequency.
Therefore, the feedback loop with a strong negative feedback serves as a
\textquotedblleft trap\textquotedblright\ of the absorption strength:\ once a
strong negative feedback $\mathbb{H}^{\prime}(h_{\alpha}^{(\mathrm{ss})}%
)\ll-1$ is formed at a certain pump frequency $\omega$, further change of the
pump frequency over a wide range does not appreciably change the detuning from
the value $\omega_{eh}^{0}+h_{\alpha}^{(\mathrm{ss})}-\omega$ and hence does
not change the pump absorption strength from the value $\chi(\omega_{eh}%
^{0}+h_{\alpha}^{(\mathrm{ss})}-\omega)$. The experimentally
observed\cite{XuNature09,LattaNaturePhys09,VinkNaturePhys09} locking of the
pump absorption strength to the resonance corresponds to the occurance of such
a \textquotedblleft trap\textquotedblright\ around the resonance point:
$\omega_{eh}^{0}+h_{\alpha}^{(\mathrm{ss})}-\omega=0$ (see
Sec.~\ref{SEC_EXAMPLE_NEGATIVE_FEEDBACK} for an example).

\item If $h_{\alpha}^{(\mathrm{ss})}$ is associated with a strong positive
feedback $\mathbb{H}^{\prime}(h_{\alpha}^{(\mathrm{ss})})>1$, then $h_{\alpha
}^{(\mathrm{ss})}$ is unstable, thus the corresponding pump absorption value
$\chi(\omega_{eh}^{0}+h_{\alpha}^{(\mathrm{ss})}-\omega)$ will not be observed
experimentally in steady state, i.e., the pump absorption value $\chi
(\omega_{eh}^{0}+h_{\alpha}^{(\mathrm{ss})}-\omega)$ is avoided. The
experimentally observed\cite{HogelePRL2012} avoidance of the pump absorption
strength from the resonance corresponds to the occurance of such a unstable
feedback on the resonance point:\ $\omega_{eh}^{0}+h_{\alpha}^{(\mathrm{ss}%
)}-\omega=0$ (see Sec.~\ref{SEC_EXAMPLE_POSITIVE_FEEDBACK} for an example).

\item If $h_{\alpha}^{(\mathrm{ss})}$ is associated with a stable, positive
feedback $0<\mathbb{H}^{\prime}(h_{\alpha}^{(\mathrm{ss})})<1$, then the
nuclear spin induced shift of the electron or hole excitation energy
$\delta\omega_{eh}$ has an opposite sign to the pump frequency change
$\delta\omega$, so that $|\delta\Delta|>|\delta\omega|$. Therefore, if a
stable, positive feedback $\mathbb{H}^{\prime}(h_{\alpha}^{(\mathrm{ss}%
)})\approx1$ is formed at a certain pump frequency $\omega$, then even a small
change of the pump frequency will lead to drastic change of the nuclear field,
which in turn shifts the detuning far away from the expected value
$\omega_{eh}^{0}+h_{\alpha}^{(\mathrm{ss})}-\omega$, corresponding to rapid
pushing of the pump absorption strength away from the value $\chi(\omega
_{eh}^{0}+h_{\alpha}^{(\mathrm{ss})}-\omega)$ (see
Sec.~\ref{SEC_EXAMPLE_WEAKPOSITIVE_FEEDBACK} for an example).
\end{itemize}

\subsection{Back action from fluctuating nuclear field: suppression or
amplification of nuclear field fluctuation}

\label{SEC_FULL_FEEDBACK}

Here we close the feedback loop by fully incorporating the back action of the
fluctuating nuclear field $\hat{h}_{z}$. In this case, the dynamics of
different nuclear spins are coupled to $\hat{h}_{z}$ through the $\hat{h}_{z}%
$-dependent transition rates $W_{\pm}(\hat{h}_{z})$. This enables the feedback
loop to control the nuclear field, both its average value and its fluctuation.
For the paradigmatic central spin model consisting of a confined electron spin
coupled to the nuclear spins through the contact hyperfine interaction
\begin{equation}
\hat{V}_{eN}=\sum_{j}a_{j,e}\mathbf{\hat{S}}_{e}\cdot\mathbf{\hat{I}}_{j},
\label{CONTACTHYPERFINE}%
\end{equation}
we identify $\hat{F}_{z}\equiv\hat{S}_{e,z}$ and the nuclear field $\hat
{h}_{z}\equiv\sum_{j}a_{j,e}\hat{I}_{j,z}$, whose strong fluctuation leads to
the detrimental effect of rapid electron spin
decoherence.\cite{KhaetskiiPRL02,MerkulovPRB02,SemenovPRB03,WitzelPRB05,WitzelPRB06,DengPRB06,YaoPRB06,YaoPRL07,LiuNJP07,YangPRB08,YangPRB08a,YangPRB09,CywinskiPRL09}
Suppressing the nuclear field fluctuation is a major direction of recent
research in spin based quantum computation.

The diagonal nuclear spin density matrix $\hat{P}(t)$ contains the information
for the population of every nuclear spin, but it is difficult to obtain such
microscopic details by solving Eq.~(\ref{RATEEQ}), even in the steady state,
because different nuclear spins are coupled to the fluctuating nuclear field
$\hat{h}_{z}$. Fortunately, the quantity of importance is the nuclear field
$\hat{h}_{z}=h_{\max}\hat{s}$. Therefore, the key is to single out the
dynamics of the nuclear field from Eq.~(\ref{RATEEQ}), as motivated by the
stochastic\cite{RudnerPRL07a,RudnerNanotechnol10,DanonPRL08,DanonPRL09,VinkNaturePhys09}
and rate equation\cite{GreilichScience07} approaches. For this purpose, we
define the probability distribution function $p(s,t)\equiv\operatorname*{Tr}%
\delta_{\hat{s},s}\hat{P}(t)$ of $\hat{s}$, i.e., the probability for the
nuclear field $\hat{s}$ to be equal to $s$ at time $t$. From Eq.~(\ref{RATEEQ}%
), we can approximately derive (see Appendix \ref{APPENDIX_FP} for details) a
closed equation of motion, i.e., the Fokker-Planck equation for $p(s,t)$:%
\begin{equation}
\frac{\partial}{\partial t}p(s,t)=\frac{\partial}{\partial s}\left[
\frac{\partial}{\partial s}D(s)p(s,t)-v(s)p(s,t)\right]  ,
\label{FOKKER_PLANCK}%
\end{equation}
where%
\begin{equation}
D(s)\approx\frac{1}{2NI}\Gamma_{\mathrm{tot}}(h_{\max}s)\left(  \frac
{2(I+1)}{3}-ss_{0}^{(1/2)}(h_{\max}s)\right)  \label{DIFFUSION}%
\end{equation}
is the diffusion coefficient and%
\begin{equation}
v(s)\approx-\Gamma_{\mathrm{tot}}(h_{\max}s)[s-s_{0}(h_{\max}s)] \label{DRIFT}%
\end{equation}
is the drift coefficient. The steady-state solution is given by%
\begin{equation}
p^{(\mathrm{ss})}(s)=\frac{D(s^{\ast})}{D(s)}p^{(\mathrm{ss})}(s^{\ast}%
)\exp\left(  \int_{s^{\ast}}^{s}\frac{v(s^{\prime})}{D(s^{\prime})}ds^{\prime
}\right)  , \label{STEADY_PS}%
\end{equation}
where $s^{\ast}$ is an arbitrary constant. The steady-state distribution
function $p^{(\mathrm{ss})}(s)$ contains all the information for the nuclear
field. Each peak of $p^{(\mathrm{ss})}(s)$ corresponds to a macroscopic
nuclear spin state (distinguished by subscript $\alpha$): the position
$s_{\alpha}^{(\mathrm{ss})}$ of the $\alpha$-th peak gives the average nuclear
field $s_{\alpha}^{(\mathrm{ss})}$, while the width $\sigma_{\alpha}$ of the
$\alpha$-th peak quantifies the fluctuation of the nuclear field $\hat{s}$
around its average value $s_{\alpha}^{(\mathrm{ss})}$.

Equations~(\ref{FOKKER_PLANCK})-(\ref{STEADY_PS}) justify and unify the
stochastic
approach\cite{RudnerPRL07a,RudnerNanotechnol10,DanonPRL08,DanonPRL09,VinkNaturePhys09}
and the rate equation approach\cite{GreilichScience07} and generalize them to
include nuclei with spins higher than 1/2. For nuclear spin-1/2, the drift
coefficient $v(s)$, the diffusion coefficient $D(s)$, and the exponent in
Eq.~(\ref{STEADY_PS}) coincide with the stochastic
approach,\cite{RudnerPRL07a,RudnerNanotechnol10,DanonPRL08,DanonPRL09,VinkNaturePhys09}
while the factor $D(s^{\ast})/D(s)$ coincides with the rate equation
approach.\cite{GreilichScience07} The exponent is associated with the
nonlinear drift $v(s)$ of the nuclear field, while the factor $D(s^{\ast
})/D(s)$ is associated with the nonlinear diffusion $D(s)$ of the nuclear
field. They correspond to two distinct feedback processes controlling the
nuclear field. As shown below, the feedback originating from the nonlinear
drift (hereafter referred to as \textquotedblleft drift\textquotedblright%
\ feedback) vanishes when the nuclear spin flip has no preferential direction
(i.e., $W_{+}=W_{-}$). By contrast, the feedback originating from the
nonlinera diffusion (hereafter referred to as \textquotedblleft
diffusion\textquotedblright\ feedback) remains efficient even for nuclear spin
flip with no preferential direction.

The rest of this subsection is organized as follows. First, we focus on
quantifying the \textquotedblleft drift\textquotedblright\ feedback by the
nuclear field feedback function. Second, we briefly discuss the
\textquotedblleft diffusion\textquotedblright\ feedback. Finally, with the
estimate of the efficiency of the \textquotedblleft drift\textquotedblright%
\ feedback and \textquotedblleft diffusion\textquotedblright\ feedback, we
conclude that the feedback is capable of recovering the intrinsic electron
spin coherence time.

\subsubsection{\textquotedblleft Drift\textquotedblright\ feedback}

The \textquotedblleft drift\textquotedblright\ feedback associated with the
exponent of $p^{(\mathrm{ss})}(s)$ has been discussed by the stochastic
approach\cite{RudnerPRL07a,RudnerNanotechnol10,DanonPRL08,DanonPRL09,VinkNaturePhys09}
for nuclear spin-1/2's. Here we focus on quantifying the control over the
nuclear field $\hat{h}_{z}=h_{\max}\hat{s}$ by the \textquotedblleft
drift\textquotedblright\ feedback with our nuclear field feedback function for
nuclei with a general spin.

Without the factor $D(s^{\ast})/D(s)$, the extremum of $p^{(\mathrm{ss})}(s)$
is determined by $v(s)=0$ as $s=s_{0}(h_{\max}s)$, equivalent to the
self-consistent equation $h=h_{\max}s_{0}(h)=\mathbb{H}(h)$ for the average
nuclear field since $h=h_{\max}s$. Further, the condition $v^{\prime
}(s_{\alpha}^{(\mathrm{ss})})\equiv\lbrack dv(s)/ds]_{s=s_{\alpha
}^{(\mathrm{ss})}}<0$ for an extremum at $s_{\alpha}^{(\mathrm{ss})}$ to be a
peak gives $[ds_{0}(h_{\max}s)/ds]_{s=s_{\alpha}^{(\mathrm{ss})}}<1$,
equivalent to the stability condition $\mathbb{H}^{\prime}(h_{\alpha
}^{(\mathrm{ss})})<1$ since $h_{\alpha}^{(\mathrm{ss})}=h_{\max}s_{\alpha
}^{(\mathrm{ss})}$. Therefore, the conditions determining the average nuclear
field and its stability are exactly the same as the mean-field treatment
discussed in Sec.~\ref{SEC_MEANFIELD_FEEDBACK}, where the nuclear field
feedback function provides a complete description. According to the analysis
there, $h=\mathbb{H}(h)$ may have multiple stable solutions $\{h_{\alpha
}^{(\mathrm{ss})}\}$, corresponding to multiple peaks of $p^{(\mathrm{ss}%
)}(s)$ at $\{s_{\alpha}^{(\mathrm{ss})}\equiv h_{\alpha}^{(\mathrm{ss}%
)}/h_{\max}\}$ and hence multiple macroscopic nuclear spin states.

The key quantity of interest is the fluctuation of the nuclear field $\hat
{s}=\hat{h}_{z}/h_{\max}$ in each macroscopic state, as quantified by the
width of the probability distribution $p^{(\mathrm{ss})}(s)$ around each peak.
For the fluctuation of the nuclear field around its average value $s_{\alpha
}^{(\mathrm{ss})}$ in the $\alpha$-th macroscopic state, we follow the
stochastic
approach\cite{RudnerPRL07a,RudnerNanotechnol10,DanonPRL08,DanonPRL09,VinkNaturePhys09}
and expand $v(s)$ around $s_{\alpha}^{(\mathrm{ss})}$ to the first order
$v(s)\approx v^{\prime}(s_{\alpha}^{(\mathrm{ss})})(s-s_{\alpha}%
^{(\mathrm{ss})})$. Then the exponential factor becomes a Gaussian peak
$\exp[-(s-s_{\alpha}^{(\mathrm{ss})})^{2}/(2\sigma_{\alpha}^{2})]$ centered at
$s_{\alpha}^{(\mathrm{ss})}$. The width of this peak is $\sigma_{\alpha
}=[D(s_{\alpha}^{(\mathrm{ss})})/|v^{\prime}(s_{\alpha}^{(\mathrm{ss}%
)})|]^{1/2},$ which, for nuclear spin-1/2's, coincides with the stochastic
approach. By substituting Eqs.~(\ref{DIFFUSION}) and (\ref{DRIFT}) into
$\sigma_{\alpha}$, we obtain%
\begin{equation}
\sigma_{\alpha}=\sigma_{\mathrm{eq}}\sqrt{\frac{1-[s_{0}^{(1/2)}(h_{\alpha
}^{(\mathrm{ss})})]^{2}}{1-\mathbb{H}^{\prime}(h_{\alpha}^{(\mathrm{ss})})}},
\label{FLUCTUATION2}%
\end{equation}
where $\sigma_{\mathrm{eq}}\equiv\lbrack(I+1)/(3NI)]^{1/2}$ is the thermal
equilibrium fluctuation of the nuclear field $\hat{s}$. Note that the
normalization $|s_{0}^{(1/2)}|\leq1$ and the stability condition
$\mathbb{H}^{\prime}(h_{\alpha}^{(\mathrm{ss})})<1$ ensures that the quantity
inside the square root of Eq.~(\ref{FLUCTUATION2}) is always finite and non-negative.

Equation (\ref{FLUCTUATION2}) shows that in the $\alpha$-th macroscopic
nuclear spin state, the nuclear field fluctuation is controlled by the nuclear
spin polarization and the feedback:

\begin{enumerate}
\item[(1)] In the absence of the e-h system, we have $s_{0}^{(1/2)}%
(h)=\mathbb{H}(h)=\mathbb{H}^{\prime}(h)=0$ and hence a unique, vanishing
nuclear field $s^{(\mathrm{ss})}=0$ in steady state. The fluctuation of the
nuclear field $\hat{s}$ is given by Eq.~(\ref{FLUCTUATION2}) as $\sigma
_{\mathrm{eq}}$, i.e., the thermal equilibrium fluctuation.

\item[(2)] If we take into account the e-h induced nuclear spin flip [wavy
arrow in Fig.~\ref{G_EHNLOOP}(b)] but neglect the back action of the nuclear
field on the e-h system [straight arrow in Fig.~\ref{G_EHNLOOP}(b)], then the
dynamics of different nuclear spins is decoupled. In steady state, each
individual nuclear spin acquires a finite polarization $s_{0}^{(1/2)}(0)$ that
suppresses its own fluctuation by a factor $\sqrt{1-[s_{0}^{(1/2)}(0)]^{2}}$.
The fluctuation of the collective nuclear field is suppressed by the same factor.

\item[(3)] Inclusion of the back action of the average nuclear field
(Sec.~\ref{SEC_MEANFIELD_FEEDBACK}) leads to multiple stable nuclear fields
$\{s_{\alpha}^{(\mathrm{ss})}\}$, so that the suppression of the nuclear field
fluctuation around $s_{\alpha}^{(\mathrm{ss})}$ becomes $\sqrt{1-[s_{0}%
^{(1/2)}(h_{\alpha}^{(\mathrm{ss})})]^{2}}$. Points (2) and (3)\ describe the
suppression of the nuclear field fluctuation by nuclear spin
polarization.\cite{CoishPRB04}

\item[(4)] If we fully take into account the back action of the fluctuating
nuclear field, then in addition to the factor $\sqrt{1-[s_{0}^{(1/2)}%
(h_{\alpha}^{(\mathrm{ss})})]^{2}}$ associated with the polarization
$s_{0}^{(1/2)}(h_{\alpha}^{(\mathrm{ss})})$ of each individual nuclear spin,
the nuclear field fluctuation is further controlled by the feedback, as
quantified by the factor $[1-\mathbb{H}^{\prime}(h_{\alpha}^{(\mathrm{ss}%
)})]^{-1/2}$ in Eq.~(\ref{FLUCTUATION2}). It can either suppress [for negative
feedback $\mathbb{H}^{\prime}(h_{\alpha}^{(\mathrm{ss})})<0$] or amplify [for
positive feedback $0<\mathbb{H}^{\prime}(h_{\alpha}^{(\mathrm{ss})})<1$] the
nuclear field fluctuation without changing the fluctuation of each individual
nuclear spin. This quantifies a previous qualitative argument\cite{XuNature09}
of feedback induced suppression of the nuclear field fluctuation: when the
fluctuation increases (decreases) the nuclear field above (below) its
macroscopic value, the negative feedback decreases (increases) the nuclear
field and tends to restore its macroscopic value.
\end{enumerate}

\subsubsection{\textquotedblleft Diffusion\textquotedblright\ feedback}

The \textquotedblleft drift\textquotedblright\ feedback is associated with the
peaks $\{s_{\alpha}^{(\mathrm{ss})}\}$ of $p^{(\mathrm{ss})}(s)$ originating
from its exponent. One distinguishing feature of the \textquotedblleft
drift\textquotedblright\ feedback is that it vanishes when the nuclear spin
flip has no preferential direction, i.e., when $W_{+}(h)=W_{-}(h)$. This is
because $W_{+}(h)=W_{-}(h)$ leads to vanishing nuclear field feedback function
$\mathbb{H}(h)\propto s(h)=0$. Consequently, the self-consistent condition
$h=\mathbb{H}(h)$ gives a unique, vanishing steady-state nuclear field and
vanishing control over the nuclear field fluctuation, so that $\sigma
=\sigma_{\mathrm{eq}}$ [Eq.~(\ref{FLUCTUATION2})]. Correspondingly, the
exponent reduces to the thermal equilibrium distribution $\exp[-s^{2}%
/(2\sigma_{\mathrm{eq}}^{2})]$.

By contrast, the \textquotedblleft diffusion\textquotedblright\ feedback is
associated with the peaks $\{\bar{s}_{\alpha}^{(\mathrm{ss})}\}$ of
$p^{(\mathrm{ss})}(s)$ originating from the sharp dips of $D(s)$. Therefore,
it does not vanish even for nuclear spin flip with no preferential direction.
Again, the peak of $p^{(\mathrm{ss})}(s)$ at $\bar{s}_{\alpha}^{(\mathrm{ss}%
)}$ corresponds to a macroscopic nuclear spin state with average nuclear field
$\bar{s}_{\alpha}^{(\mathrm{ss})}$. The fluctuation $\bar{\sigma}_{\alpha}$ of
the nuclear field around the average value $\bar{s}_{\alpha}^{(\mathrm{ss})}$
is determined by the width of the peak. For example, the nuclei induced
frequency focusing observed by Greilich \textit{et al.}%
\cite{GreilichScience07} upon periodic pulsed excitation of the electron spin
originates from such \textquotedblleft diffusion\textquotedblright\ feedback.
There, $D(s)$ exhibits multiple sharp dips spaced by $2\pi\nu_{\mathrm{rep}%
}/h_{\max}$ as determined by the pulse repetition rate $\nu_{\mathrm{rep}}$.
Recently, based on the electron-induced nuclear spin flip with no preferential
direction, Issler \textit{et al.}\cite{IsslerPRL10} proposed a nuclear spin
cooling scheme with nuclear field selective coherent population trapping,
where the suppression of the nuclear field fluctuation was analyzed with Monte
Carlo simulation. This scheme is an excellent example of the \textquotedblleft
diffusion\textquotedblright\ feedback: the coherent dark state dip of the
electron population introduces a sharp dip into the electron-induced nuclear
spin flip rates $W_{\pm}(h_{\max}s)$ and hence the diffusion coefficient
$D(s)$. Consequently, the distribution function $p^{(\mathrm{ss})}(s)$
exhibits a narrow peak, corresponding to a finite nuclear field with
suppressed fluctuation.

\subsubsection{Recovering intrinsic electron spin coherence time by feedback}

First we estimate the efficiency of the \textquotedblleft
drift\textquotedblright\ feedback and the \textquotedblleft
diffusion\textquotedblright\ feedback. Suppose that the characteristic scale
for the nuclear spin transition rates $W_{\pm}(h)$ to change appreciably is
$\delta h$. For the \textquotedblleft drift\textquotedblright\ feedback, the
maximal feedback strength is roughly estimated as $|\mathbb{H}^{\prime
}(h)|\sim h_{\max}/\delta h$, where we have assumed that the maximal
achievable nuclear spin polarization $\sim O(1)$. Therefore, according to
Eq.~(\ref{FLUCTUATION2}), the typical fluctuation of the nuclear field
$\hat{h}_{z}$ under the \textquotedblleft drift\textquotedblright\ feedback
is
\[
h_{\max}\sigma\sim\sqrt{a_{z}\delta h}.
\]
On the other hand, the typical width $\bar{\sigma}$ of a dip of $D(s)$ is
given by the characteristic scale for $D(s)$ to change, i.e., $\bar{\sigma
}\sim\delta h/h_{\max}$, thus the typical fluctuation of the nuclear field
$\hat{h}_{z}$ under the \textquotedblleft diffusion\textquotedblright%
\ feedback is
\[
h_{\max}\bar{\sigma}\sim\delta h.
\]
Note that $h_{\max}\sigma\propto\sqrt{\delta h}$ and $h_{\max}\bar{\sigma
}\propto\delta h$ scales differently with $\delta h$.

Second we compare the efficiency of the \textquotedblleft
drift\textquotedblright\ feedback with the \textquotedblleft
diffusion\textquotedblright\ feedback. If $\delta h\ll a_{z}$, i.e., the
nuclear spin flip rates $W_{\pm}(h)$ change drastically upon a slight change
of the nuclear field induced by a single nuclear spin flip event, then
$h_{\max}\bar{\sigma}\ll h_{\max}\sigma\ll a_{z}$, i.e., the \textquotedblleft
diffusion\textquotedblright\ feedback is more efficient. In this case, the
rate of the electron spin decoherence due to the nuclear field fluctuation is
much smaller than $a_{z}$. In the opposite case $\delta h\gg a_{z}$, we have
$h_{\max}\bar{\sigma}\gg h_{\max}\sigma\gg a_{z}$, i.e., the \textquotedblleft
drift\textquotedblright\ feedback is more efficient. In this case, the rate of
the electron spin decoherence due to nuclear field fluctuation is much larger
than $a_{z}$.

Since $W_{\pm}(h)$ are determined by the e-h fluctuation, the typical scale
$\delta h$ for $W_{\pm}(h)$ to change appreciably is the relevant e-h
relaxation rate $\gamma_{eh}$. Typically the orbital relaxation of the e-h
system is much faster than their spin relaxation, thus the smallest
$\gamma_{eh}$ corresponds to the \textquotedblleft intrinsic\textquotedblright%
\ electron or hole spin relaxation rate $1/T_{2,e}$ or $1/T_{2,h}$. Therefore,
as long as the limit $\delta h\sim1/T_{2,e}$ is achieved, the
\textquotedblleft diffusion\textquotedblright\ feedback can suppress the
nuclear field fluctuation to $h_{\max}\bar{\sigma}\sim1/T_{2,e}$ and hence
recover the intrinsic electron spin coherence time $T_{2,e}$. On the other
hand, if $a_{z}\ll1/T_{2,e}$, then achievement of $\delta h\sim1/T_{2,e}$ also
enables the \textquotedblleft drift\textquotedblright\ feedback to suppress
the nuclear field fluctuation to $h_{\max}\sigma\sim\sqrt{a_{z}/T_{2,e}}%
\ll1/T_{2,e}$ and hence recover $T_{2,e}$.

\section{Example: nuclear spin dynamics through non-collinear dipolar
hyperfine interaction}

To exemplify our general theory, we consider the electron-hole-nuclei feedback
loop in Fig.~\ref{G_ENLOOP_HNLOOP}(b). It was first proposed by Xu \textit{et
al.}\cite{XuNature09} to explain the experimentally observed symmetric locking
of the pump absorption strength and suppressed nuclear field fluctuation, and
the key element of this loop, i.e., the mechanism of hole-driven dynamic
nuclear polarization, was established recently.\cite{YangPRB2012} While Ref.
\onlinecite{YangPRB2012} introduced the concept of the feedback loop for this
single spin with spin bath problem, the current work differs from it in the
perspective of a general theory providing key insight into the important
consequences. Here, instead of explicitly classifying the density matrix
elements into the \textquotedblleft slow\textquotedblright\ ones and the
\textquotedblleft fast\textquotedblright\ ones (which is rather tedious), we
directly apply the general result Eq. (\ref{WPN}) to the electron-hole-nuclei
feedback loop and utilize the quantum regression theorem for a compact
derivation. This feedback loop was also utilized by Ladd \textit{et
al.}\cite{LaddPRL10,LaddSPIE2011} to explain the experimentally observed
hysteretic sawtooth pattern in the electron spin free induction decay. An
advantage of exemplifying our theory with this feedback loop instead of the
more intensively investigated electron-nuclei feedback loop
[Fig.~\ref{G_ENLOOP_HNLOOP}(a)], is that this loop can realize all the
interesting regimes discussed in our general theory, i.e., bistability, strong
negative feedback, and positive feedback.

The essential difference between the electron-hole-nuclei feedback loop
[Fig.~\ref{G_ENLOOP_HNLOOP}(b)] and the electron-nuclei feedback loop
[Fig.~\ref{G_ENLOOP_HNLOOP}(a)] is that in Fig.~\ref{G_ENLOOP_HNLOOP}(b), the
nuclear spins are flipped by the non-collinear interaction $\propto\hat
{S}_{h,z}(\hat{I}_{+}+\hat{I}_{-})$ with the hole, while in
Fig.~\ref{G_ENLOOP_HNLOOP}(a), the nuclear spins are flipped by the contact
hyperfine interaction $\propto\hat{S}_{e,+}\hat{I}_{-}+\hat{S}_{e,-}\hat
{I}_{+}$ with the electron. The former process is not accompanied by the hole
spin flip, so it involves a very small energy mismatch ($\sim$ nuclear Zeeman
splitting) and hence is nearly resonant. By contrast, the latter process is
accompanied by the electron spin flip, so it involves a much larger energy
mismatch ($\sim$ electron Zeeman splitting) and hence is off-resonant.
Consequently, although the hole-nuclear non-collinear interaction is much
weaker than the electron-nuclear contact hyperfine interaction, the tremendous
resonant enhancement originating from a small energy mismatch could make the
strength of the former process comparable with the latter
process.\cite{YangPRB2012} Recently, this mechanism is generalized to the case
of non-collinear electron-nuclear interaction (which arises from nuclear
quadrupolar effect \cite{HuangPRB2010}) to explain the experimentally observed
avoidance of resonant absorption \cite{LattaPRL2011}. Under optical excitation
conditions, both the electron-nuclear and hole-nuclear non-collinear hyperfine
interaction may play a role in determining the nuclear polarization. However,
the relative contributions from the electron and the hole remains an open
issue.\cite{UrbaszekRMP2013}

For the realistic physical system corresponding to the electron-hole-nuclei
feedback loop [Fig.~\ref{G_ENLOOP_HNLOOP}(b)], we consider a negatively
charged QD subjected to an external magnetic field $B$ along the QD growth
direction (defined as the $z$ axis). A right circularly polarized continuous
wave laser applied in the Faraday configuration couples the spin-up electron
level $\left\vert 0\right\rangle $ to the spin-up trion level $\left\vert
1\right\rangle $. The spin-up trion consists of two inert electrons in the
spin singlet and one unpaired spin-up hole. Since the hole is the only active
member of the trion, hereafter we refer to the trion as hole for brevity. The
electron level $\left\vert 0\right\rangle $ and the hole level $\left\vert
1\right\rangle $ form the e-h system illustrated in Fig.~\ref{G_ENLOOP_HNLOOP}%
(b). The optically pumped e-h system is described by the Hamiltonian
\[
\hat{H}_{eh}(t)=-\omega_{0}\hat{\sigma}_{00}+\frac{\Omega_{R}}{2}\left(
\hat{\sigma}_{10}e^{-i\omega t}+\hat{\sigma}_{01}e^{i\omega t}\right)
+\hat{H}_{\mathrm{damp}},
\]
where $\omega_{0}$ is the \textquotedblleft bare\textquotedblright\ e-h
excitation energy in the absence of the nuclear spins, $\hat{\sigma}%
_{ji}\equiv\left\vert j\right\rangle \left\langle i\right\vert $, $\omega$ is
the laser frequency$,$ $\hat{H}_{\mathrm{damp}}$ denotes the coupling to the
vacuum electromagnetic fluctuation that leads to spontaneous emission
$\left\vert 1\right\rangle \rightarrow\left\vert 0\right\rangle $ with rate
$\gamma_{1}$ and hole dephasing with total rate $\gamma_{2}$ in the Lindblad
form, and $\Omega_{R}=-e\mathbf{E}_{0}\cdot\left\langle 1\right\vert
\mathbf{r}\left\vert 0\right\rangle $ is the Rabi frequency: the coupling
between the electric dipole $-e\left\langle 1\right\vert \mathbf{r}\left\vert
0\right\rangle $ and the pump electric field $\mathbf{E}(t)=\mathbf{E}_{0}%
\cos\omega t$. The coupling between the e-h system and the nuclear spins,
after being projected into the relevant hilbert space spanned by $\{\left\vert
0\right\rangle $,$\left\vert 1\right\rangle \}$, is%
\[
\hat{V}=\frac{1}{2}\hat{\sigma}_{00}\sum_{j}a_{j,e}\hat{I}_{j,z}+\hat{\sigma
}_{11}\sum_{j}\tilde{a}_{j,h}(\hat{I}_{j,+}+\hat{I}_{j,-}),
\]
where the first term is the diagonal part of the electron-nuclear contact
hyperfine interaction [leading to the nuclear field back action, as denoted by
the straight arrow in Fig.~\ref{G_ENLOOP_HNLOOP}(b)] and the second term is
the non-collinear part of the hole-nuclear dipolar hyperfine interaction
[leading to nuclear spin flip, as denoted by the wavy arrow in
Fig.~\ref{G_ENLOOP_HNLOOP}(b)]. A brief summary of the nuclear spin dynamics
driven by other parts of the electron-nuclear and hole-nuclear interactions
could be found elsewhere.\cite{YangPRB2012} The total Hamiltonian $\hat
{H}(t)=\hat{H}_{N}+\hat{H}_{eh}(t)+\hat{V}$ assumes the same form as our
theory [Eq.~(\ref{HAMILTONIAN})], where $\hat{H}_{N}$ is the nuclear spin
Hamiltonian given in Eq.~(\ref{HN}).

To apply our theory to this model, we further put the coupling $\hat{V}$ into
the same form as our theory [Eq.~(\ref{V})] by identifying $\hat{F}%
_{z}(t)=\hat{\sigma}_{00}$, $\hat{F}_{\pm}(t)=\hat{\sigma}_{11}$,
$a_{j,z}=a_{j,e}/2$, and $a_{j,\pm}=\tilde{a}_{j,h}$. Then, according to our
theory, after adiabatically eliminating the e-h dynamics, the diagonal part
$\hat{P}(t)$ of the nuclear spin density matrix obeys Eq.~(\ref{RATEEQ}),
where the transition rate $W_{j,\pm}(\hat{h}_{z})$ is equal to Eq.~(\ref{WPN})
plus $\Gamma_{1,j}/2$, which accounts for the nuclear spin depolarization due
to other nuclear spin relaxation mechanisms.

As in the general theory, we consider identical nuclear spins ($I_{j}=I$,
$\omega_{j,N}=\omega_{N}$, and $\Gamma_{1,j}=\Gamma_{1}$) uniformly coupled to
the electron and the hole ($a_{j,e}=a_{e}$, $\tilde{a}_{j,h}=\tilde{a}_{h}$).
For a typical self-assembled QD containing $N=10^{4}$ nuclear spins subjected
to an external magnetic field $B\sim1$ T, the order of magnitude of relevant
parameters\cite{LiuNJP07,XuNature09,LattaNaturePhys09,YangPRB2012} is listed
in Table \ref{TableKey}.%

%TCIMACRO{\TeXButton{B}{\begin{table}[tbp] \centering}}%
%BeginExpansion
\begin{table}[tbp] \centering
%EndExpansion
\caption{Order of magnitude of relevant parameters (unit: $\mathrm{ns}^{-1}$, with the convention $\hbar=1$ understood) for a
typical self-ensembled QD containing $N=10^{4}$ nuclear spins under a magnetic field $B\sim$ 1 T.}%
\begin{tabular*}
{\columnwidth}[c]{@{\extracolsep{\fill}}ccccccc}\hline
$\Omega_{R}$ & $\gamma_{1}$ & $\gamma_{2}$ & $\left\vert \omega_{N}\right\vert
$ & $a_{e}$ & $\left\vert \tilde{a}_{h}\right\vert $ & $\Gamma_{1}%
$\\\hline\hline
$1$ & $1$ & $1$ & $0.1$ & $10^{-2}$ & $10^{-5}$ & $10^{-10}$\\\hline
\end{tabular*}
\label{TableKey}%
%TCIMACRO{\TeXButton{E}{\end{table}}}%
%BeginExpansion
\end{table}%
%EndExpansion

For the present model, the nuclear spin transition rates $W_{\pm}(h)$ and
hence the diffusion coefficient $D(s)$ do not exhibit sharp dips, so the
\textquotedblleft diffusion\textquotedblright\ feedback is negligible. In the
rest, we only consider the \textquotedblleft drift\textquotedblright%
\ feedback. First we calculate the nuclear field feedback function
$\mathbb{H}(h)$. Then we use it to quantify the \textquotedblleft
drift\textquotedblright\ feedback (including absorption strength locking or
avoidance and the suppression or amplification of the nuclear field
fluctuation) for three cases: strong negative feedback, strong positive
feedback, and weak positive feedback.

\subsection{Nuclear field feedback function}

Following the theory in Sec.~\ref{SEC_CONSTANT_FEEDBACK}, to obtain the
nuclear field feedback function $\mathbb{H}(h)$, we need the nuclear spin flip
rates $W_{\pm}(h)=\tilde{a}_{h}^{2}C(h,\mp\omega_{N})+\Gamma_{1}/2$, where
$\hat{h}_{z}$ has been replaced by a constant $h$ and, from Eq.~(\ref{WPN}),
\[
C(h,\nu)\equiv\int_{-\infty}^{\infty}e^{i\nu t}dt\ \operatorname*{Tr}%
\nolimits_{eh}\hat{\sigma}_{11}^{\mathrm{I}}(h,t)\hat{\sigma}_{11}%
^{\mathrm{I}}(h,0)\hat{\rho}_{eh}^{(\mathrm{ss})}(h,0).
\]
We note that for the present model, the $h$-dependent e-h evolution $\hat
{U}_{eh}(h,t)$ [see Eq.~(\ref{UEH_EFF})] is obtained from the free e-h
evolution $\hat{U}_{eh}(t)$ [see Eq.~(\ref{UEH})] by replacing the
\textquotedblleft bare\textquotedblright\ e-h excitation energy $\omega_{0}$
with the actual e-h excitation energy $\omega_{0}-h$ or equivalently, by
replacing the nominal detuning $\Delta_{0}\equiv\omega_{0}-\omega$ with the
actual detuning $\Delta\equiv\Delta_{0}-h$. So $C(h,\nu)$ is obtained from
$C(h=0,\nu)$ by replacing $\Delta_{0}$ with $\Delta$, i.e., $C(h,\nu)$ is a
function of $\Delta$ and $\nu$. Therefore, the $h$-dependence of $C(h,\nu)$
and hence other $h$-dependent quantities such as $s_{0}(h)$ and $\mathbb{H}%
(h)\equiv h_{\max}s_{0}(h)$ entirely comes from their dependence on $\Delta$.
To emphasize this dependence, we use functions of $\Delta$ for functions of
$h$, e.g., $C(\Delta,\nu)$ for $C(h,\nu)$, $s_{0}(\Delta)$ for $s_{0}(h)$, and
$\mathbb{H}(\Delta)$ for $\mathbb{H}(h)$, etc.

In the absence of the nuclear spin depolarization $(\Gamma_{1}=0$),
Eq.~(\ref{S0}) gives the steady-state nuclear spin polarization
\[
s_{0}^{(1/2\mathrm{)}}(\Delta)=\frac{C(\Delta,-\omega_{N})-C(\Delta,\omega
_{N})}{C(\Delta,-\omega_{N})+C(\Delta,\omega_{N})}%
\]
for nuclear spin-1/2's. For nuclear spin-$I$'s, the steady-state nuclear spin
polarization $s_{0}(\Delta)$ is obtained from Eq.~(\ref{S0I}) or $s_{0}%
(\Delta)\approx\lbrack2(I+1)/3]s_{0}^{(1/2\mathrm{)}}(\Delta)$ for weak
polarization $\left\vert s_{0}(\Delta)\right\vert \ll1$. In the presence of
nuclear spin depolarization, $s_{0}^{(1/2\mathrm{)}}(\Delta)$ is reduced by a
factor $1+\Gamma_{1}/\Gamma_{p}(\Delta)$ determined by the ratio between the
hole-induced nuclear spin flip rate $\Gamma_{p}(\Delta)=\tilde{a}_{h}%
^{2}[C(\Delta,-\omega_{N})+C(\Delta,\omega_{N})]$ and the nuclear spin
depolarization rate $\Gamma_{1}$. For $\Gamma_{1}=0$,\ by evaluating
$C(\Delta,\nu)$ (see Appendix \ref{APPENDIX_QR}) through the quantum
regression theorem,\cite{ScullyQuantumOptics} we obtain explicit analytical
experessions
\begin{align}
\Gamma_{p}(\Delta)  &  \approx\frac{4\tilde{a}_{h}^{2}\gamma_{1}W(\Delta
)}{[\gamma_{1}+2W(\Delta)]^{3}}c_{1}(\Delta),\label{GAMMAP}\\
s_{0}^{(1/2)}(\Delta)  &  \approx-\frac{\Delta\omega_{N}}{\Delta^{2}%
+\gamma_{2}^{2}}\frac{\gamma_{1}}{\gamma_{2}}\frac{c_{0}(\Delta)}{c_{1}%
(\Delta)}, \label{S012}%
\end{align}
up to leading order of the small quantity $\varepsilon\equiv\omega_{N}%
/\gamma_{1,2}$, where $c_{0}(\Delta)=\gamma_{2}/\gamma_{1}+1/2+f(\Delta
)+W(\Delta)/\gamma_{1}$ and $c_{1}(\Delta)=1+[\gamma_{1}/(2\gamma
_{2})]f(\Delta)+W(\Delta)/\gamma_{1}$ are non-negative constants (because
$\gamma_{2}\geq\gamma_{1}/2$) with $f(\Delta)\equiv(\gamma_{2}^{2}-\Delta
^{2})/(\gamma_{2}^{2}+\Delta^{2})$ and $W(\Delta)=2\pi(\Omega_{R}/2)^{2}%
\delta^{(\gamma_{2})}(\Delta)$ is the optical pumping rate from level
$\left\vert 0\right\rangle $ to level $\left\vert 1\right\rangle $, with
$\delta^{(\gamma_{2})}(\Delta)\equiv$ $(\gamma_{2}/\pi)/(\Delta^{2}+\gamma
_{2}^{2})$ the energy-conserving $\delta$-function broadened by hole
dephasing. Near resonance $\Delta=0$, we estimate $\Gamma_{p}\sim\tilde{a}%
_{h}^{2}/\gamma_{1}\sim0.1$ $\mathrm{s}^{-1}$ to be comparable with the
typical nuclear spin depolarization rate $\Gamma_{1}$ (see Table
\ref{TableKey}).

\begin{figure}[ptb]
\includegraphics[width=\columnwidth]{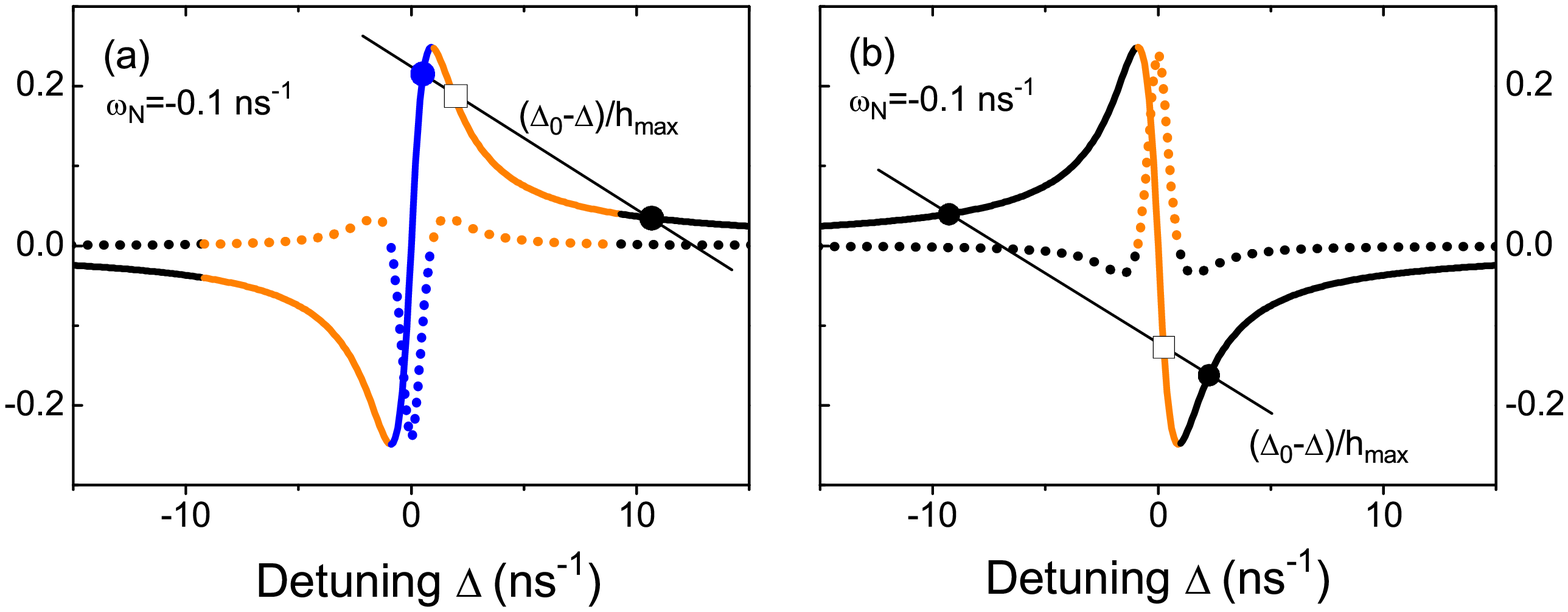}\caption{(color
online). $s_{0}(\Delta)$ (solid curves) and $\mathbb{H}^{\prime}(\Delta)/500$
(dotted curves) vs. detuning $\Delta$ for nuclear spin-9/2's (e.g., InAs QD)
for $\Gamma_{1}=0$, (a) $\omega_{N}=-0.1$ $\mathrm{ns}^{-1}$ and (b)
$\omega_{N}=0.1$ $\mathrm{ns}^{-1}$. Other parameters are from Table
\ref{TableKey}. Dark (Black and blue) segments of the curves correspond to
stable feedback $\mathbb{H}^{\prime}(\Delta)<1$. Light (Orange) segments
correspond to unstable feedback $\mathbb{H}^{\prime}(\Delta)>1$. The straight
lines denote $(\Delta_{0}-\Delta)/h_{\max}$, whose intersections with
$s_{0}(\Delta)$ give the steady-state nuclear field $\{s_{\alpha
}^{(\mathrm{ss})}\},$ including stable (filled circles) and unstable (empty
squares) ones.}%
\label{G_FEEDBACK_FUNCTION}%
\end{figure}

Equations~(\ref{GAMMAP}) and (\ref{S012}) are the key results of the recently
established mechanism of dynamic nuclear polarization by non-collinear
hyperfine interaction.\cite{YangPRB2012} In addition to the dependence
$s_{0}^{(1/2)}(\Delta),s_{0}(\Delta)\sim\Delta$ responsible for the absorption
strength locking or avoidance,\cite{YangPRB2012} another distinguishing
feature of this mechanism is that in the absence of nuclear spin
depolarization $(\Gamma_{1}=0)$, the steady-state nuclear spin polarization
$s_{0}(\Delta)$ and hence the nuclear field feedback function $\mathbb{H}%
(\Delta)=h_{\max}s_{0}(\Delta)$ still strongly depend on the optical detuning
$\Delta=\Delta_{0}-h$, so that the \textquotedblleft drift\textquotedblright%
\ feedback as quantified by the feedback strength $\mathbb{H}^{\prime}%
(\Delta)=d\mathbb{H}(\Delta)/dh=-d\mathbb{H}(\Delta)/d\Delta=-h_{\max}%
ds_{0}(\Delta)/d\Delta$ remains efficient, as shown in
Fig.~\ref{G_FEEDBACK_FUNCTION}. By contrast, for other dynamic nuclear
polarization mechanisms such as the Overhauser or reverse Overhauser
effect,\cite{OverhauserPR53,RudnerPRL07,DanonPRL08,DanonPRL09} if $\Gamma
_{1}=0$, then $s_{0}(\Delta)$ and hence $\mathbb{H}(\Delta)$ are independent
of $\Delta$, so that the \textquotedblleft drift\textquotedblright\ feedback
vanishes (as mentioned in the introduction, the stochastic approach
constructed for those mechanisms shows that suppression of the nuclear field
fluctuation comes entirely from the competition between dynamic nuclear
polarization and nuclear spin depolarization). In other words, the
\textquotedblleft drift\textquotedblright\ feedback is extrinsic to those
mechanisms but intrinsic to the dynamic nuclear polarization induced by
non-collinear hyperfine interaction. Below we focus on this intrinsic
\textquotedblleft drift\textquotedblright\ feedback by setting $\Gamma_{1}=0$.

In addition to the specific results in Fig.~\ref{G_FEEDBACK_FUNCTION}, we can
also analyze $s_{0}(\Delta)$ and $\mathbb{H}^{\prime}(\Delta)$ more generally.
First, $s_{0}^{(1/2)}(\Delta)$, $s_{0}(\Delta)$, and $\mathbb{H}^{\prime
}(\Delta)$ are reversed upon reversal of the Zeeman frequency $\omega_{N}$.
Second, by dropping the $O(1)$ factors of $(\gamma_{1}/\gamma_{2})c_{0}%
(\Delta)/c_{1}(\Delta)$ in Eq.~(\ref{GAMMAP}), we obtain the maximal magnitude
of $s_{0}^{(1/2)}(\Delta)$:
\[
\left\vert s_{0}^{(1/2)}\right\vert _{\max}\sim\frac{\left\vert \omega
_{N}\right\vert }{\gamma_{2}}.
\]
The typical magnitude is $|s_{0}^{(1/2)}|_{\max}\sim10\%$ (based on Table
\ref{TableKey}). Third, near the resonance $\left\vert \Delta\right\vert
\ll\gamma_{2}$, $s_{0}^{(1/2)}(\Delta)\propto\Delta$ is linear in $\Delta$ and
the feedback strength is maximal:
\[
\mathbb{H}^{\prime}(0)\sim I(I+1)\frac{Na_{e}}{\gamma_{2}}\frac{\omega_{N}%
}{\gamma_{2}},
\]
where we have used $s_{0}(\Delta)\sim(I+1)s_{0}^{(1/2)}(\Delta)$. Based on
Table \ref{TableKey}, the typical magnitude is $|\mathbb{H}^{\prime}%
(0)|\sim10I(I+1)$. Thus the feedback near the resonance is strongly negative
(positive) for negative (positive) nuclear Zeeman frequency $\omega_{N}$.
These results agree with Fig.~\ref{G_FEEDBACK_FUNCTION}.

\subsection{Back action from nuclear field}

We consider three cases: (i) strong negative feedback [$\omega_{N}=-0.1$,
Fig.~\ref{G_FEEDBACK_FUNCTION}(a)], (ii) strong positive feedback [$\omega
_{N}=0.1$, Fig.~\ref{G_FEEDBACK_FUNCTION}(b)], and (iii) weak positive
feedback. Case (iii) can be realized by considering nuclear spin-1/2's
(instead of nuclear spin-9/2's) with a larger hole dephasing rate $\gamma_{2}%
$. The back action of the nuclear field induces three effects: bistability,
absorption strength locking or avoidance, and suppression or amplification of
the nuclear field fluctuation. In the following, we illustrate these three
effects for each case.

\subsubsection{Strong negative feedback}

\label{SEC_EXAMPLE_NEGATIVE_FEEDBACK}

\begin{figure}[ptb]
\includegraphics[width=\columnwidth]{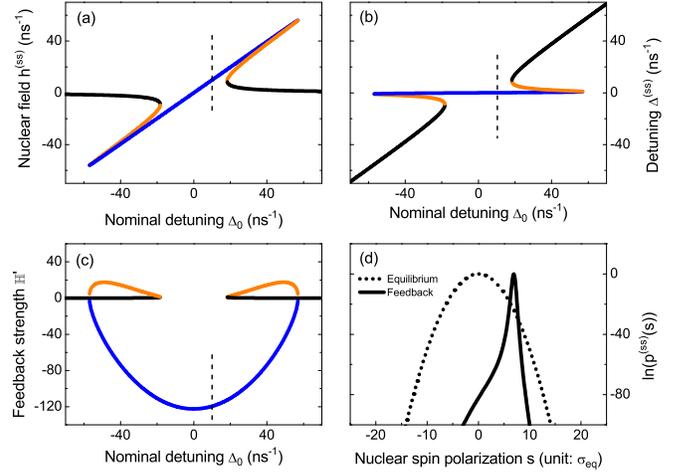}\caption{(color
online). Steady-state (a) nuclear field $h^{(\mathrm{ss})}$, (b) detuning
$\Delta^{(\mathrm{ss})}\equiv\Delta_{0}-h^{(\mathrm{ss})}$, and (c) feedback
strength $\mathbb{H}^{\prime}$ vs. nominal detuning $\Delta_{0}$ for
$\omega_{N}=-0.1$ $\mathrm{ns}^{-1}$ and $I=9/2$. Other parameters are from
Table \ref{TableKey}. Dark (Black and blue) segments of the curves correspond
to stable feedback $\mathbb{H}^{\prime}<1$. Light (Orange) segments correspond
to unstable feedback $\mathbb{H}^{\prime}>1$. (d) Solid curve: steady-state
distribution $p^{(\mathrm{ss})}(s)$ of the nuclear field $\hat{s}$ at
$\Delta_{0}=10$ $\mathrm{ns}^{-1}$ [marked by dashed straight lines in panels
(a)-(c)]. The thermal equilibrium distribution (dotted curve) is also shown
for comparison. The unit of the nuclear spin polarization is $\sigma
_{\mathrm{eq}}=[(I+1)/(3NI)]^{1/2}$, the thermal equilibrium fluctuation.}%
\label{G_FEEDBACK_NEGATIVE}%
\end{figure}

Here we consider negative nuclear Zeeman frequency $\omega_{N}=-0.1$
$\mathrm{ns}^{-1}$ [Fig.~\ref{G_FEEDBACK_FUNCTION}(a)], corresponding to a
strong negative feedback near the resonance $\Delta=0$.

For each nominal detuning $\Delta_{0}$, the steady-state nuclear field
$h^{(\mathrm{ss})}=h_{\max}s^{(\mathrm{ss})}$ is determined by the nonlinear
equation $h=\mathbb{H}(\Delta_{0}-h)$ or equivalently $s=s_{0}(\Delta
_{0}-h_{\max}s)$, i.e., $s^{(\mathrm{ss})}$ corresponds to the intersections
of $s_{0}(\Delta)$ and $(\Delta_{0}-\Delta)/h_{\max}$
[Fig.~\ref{G_FEEDBACK_FUNCTION}(a)], where $\Delta\equiv\Delta_{0}%
-h=\Delta_{0}-h_{\max}s$. For vanishing $h_{\max}$ and hence vanishing
feedback strength $\mathbb{H}^{\prime}=0$, we have a unique solution
$s^{(\mathrm{ss})}=s_{0}(\Delta_{0})$. For large $h_{\max}$ and hence strong
feedback, $(\Delta_{0}-\Delta)/h_{\max}$ becomes less steep and has up to
three intersections with $s_{0}(\Delta)$, corresponding to three steady-state
nuclear fields. The stability condition $\mathbb{H}^{\prime}<1$ gives
$d\mathbb{H}(\Delta)/d\Delta>-1$, i.e., the slope of $s_{0}(\Delta
)=\mathbb{H}(\Delta)/h_{\max}$ should be larger than that of $(\Delta
_{0}-\Delta)/h_{\max}$. So the three steady-state nuclear fields consist of
two stable ones [filled circles in Fig.~\ref{G_FEEDBACK_FUNCTION}(a)]
separated by a unstable one [empty square in Fig.~\ref{G_FEEDBACK_FUNCTION}(a)].

These steady-state nuclear fields vs. the nominal detuning $\Delta_{0}$ are
shown in Fig.~\ref{G_FEEDBACK_NEGATIVE}(a). A striking feature is that in the
middle segment (crossed by the dashed line), over a wide range of the nominal
detuning $\Delta_{0}\sim\lbrack-60$ $\mathrm{ns}^{-1},60$ $\mathrm{ns}^{-1}]$,
the stable nuclear field $h^{(\mathrm{ss})}$ follows $\Delta_{0}$ as
$h^{(\mathrm{ss})}\approx\Delta_{0}$, so that the steady-state detuning
$\Delta^{(\mathrm{ss})}=\Delta_{0}-h^{(\mathrm{ss})}$ is locked to resonance
$\Delta^{(\mathrm{ss})}\approx0$ [Fig.~\ref{G_FEEDBACK_NEGATIVE}(b)]. As
discussed in Sec.~\ref{SEC_MEANFIELD_FEEDBACK}, this behavior originates from
the strong negative feedback $\mathbb{H}^{\prime}(\Delta)\ll-1$ that occurs
near $\Delta\approx0$ [Fig.~\ref{G_FEEDBACK_FUNCTION}(a)].

The strong locking of $\Delta^{(\mathrm{ss})}$ to resonance $\Delta
^{(\mathrm{ss})}\approx0$ over $\Delta_{0}\sim\lbrack-60$ $\mathrm{ns}%
^{-1},60$ $\mathrm{ns}^{-1}]$ in turn keeps the nuclear field feedback
strength $\mathbb{H}^{\prime}(\Delta^{(\mathrm{ss})})\approx\mathbb{H}%
^{\prime}(0)$ strongly negative over the same range of $\Delta_{0}$
[Fig.~\ref{G_FEEDBACK_NEGATIVE}(c)], although strong negative feedback
$\mathbb{H}^{\prime}(\Delta)\ll-1$ only appears over $\Delta\sim\lbrack-1$
ns$^{-1},1$ ns$^{-1}]$. This strong negative feedback over a wide range of
$\Delta_{0}$ in turn strongly suppresses the nuclear field fluctuation over
the same range of $\Delta_{0}$ according to Eq.~(\ref{FLUCTUATION2}). For
example, at $\Delta_{0}=10$ $\mathrm{ns}^{-1}$ [marked by the straight dashed
line in Figs.~\ref{G_FEEDBACK_NEGATIVE}(a)-\ref{G_FEEDBACK_NEGATIVE}(c)], the
detuning is still locked to resonance $\Delta^{(\mathrm{ss})}\approx0$, so the
feedback is still strongly negative. Correspondingly, the width of the peak of
$p^{(\mathrm{ss})}(s)$ [solid curve in Fig.~\ref{G_FEEDBACK_NEGATIVE}(d)],
which quantifies the nuclear field fluctuation under the feedback control, is
much narrower than the width of the thermal distribution [dotted curve in
Fig.~\ref{G_FEEDBACK_NEGATIVE}(d)].

\subsubsection{Strong positive feedback}

\label{SEC_EXAMPLE_POSITIVE_FEEDBACK}

\begin{figure}[ptb]
\includegraphics[width=\columnwidth]{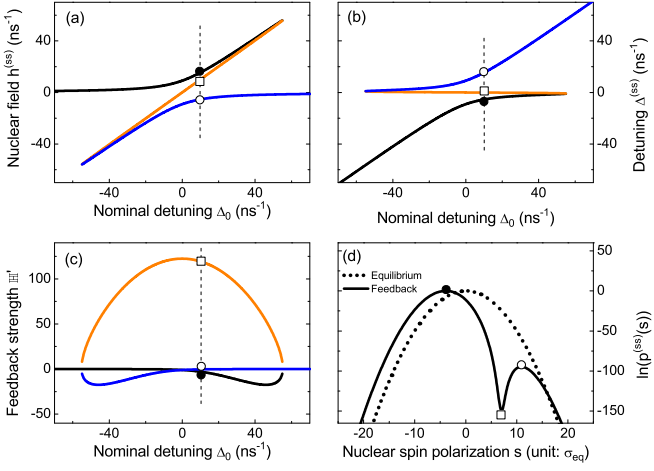}\caption{(color
online). The same as Fig.~\ref{G_FEEDBACK_NEGATIVE}, except that $\omega
_{N}=0.1$ $\mathrm{ns}^{-1}$. In panel (d), the two peaks (marked by filled
and empty circles) and the dip (marked by empty square) of $p^{(\mathrm{ss}%
)}(s)$ correspond, respectively, to the two stable solutions and the unstable
solution in panels (a)-(c) (marked by corresponding symbols).}%
\label{G_FEEDBACK_POSITIVE}%
\end{figure}

Here we consider positive nuclear Zeeman frequency $\omega_{N}=0.1$
$\mathrm{ns}^{-1}$ [Fig.~\ref{G_FEEDBACK_FUNCTION}(b)], corresponding to a
strong positive (and hence \textit{unstable}) feedback near the resonance
$\Delta=0$.

In this case, for each nominal detuning $\Delta_{0}$, $(\Delta_{0}%
-\Delta)/h_{\max}$ could also have up to three intersections with the curve
$s_{0}(\Delta)$ [Fig.~\ref{G_FEEDBACK_FUNCTION}(b)], corresponding to three
steady-state nuclear fields $\{h_{\alpha}^{(\mathrm{ss})}\}$, consisting of
two stable ones [filled circles in Fig.~\ref{G_FEEDBACK_FUNCTION}(b)]
separated by a unstable one [empty square in Fig.~\ref{G_FEEDBACK_FUNCTION}(b)].

These steady-state nuclear fields vs. the nominal detuning $\Delta_{0}$ are
shown in Fig.~\ref{G_FEEDBACK_POSITIVE}(a). As a result of the strong positive
feedback $\mathbb{H}^{\prime}(\Delta)\gg1$ near $\Delta\approx0$
[Fig.~\ref{G_FEEDBACK_FUNCTION}(b)], although the \textit{unstable} (and hence
not observable) nuclear field (marked by the empty square) follows the nominal
detuning $\Delta_{0}$ and locks the detuning $\Delta^{(\mathrm{ss})}%
=\Delta_{0}-h^{(\mathrm{ss})}$ to resonance, the two \textit{stable }nuclear
fields (marked by empty and filled circles) always push the detuning
$\Delta^{(\mathrm{ss})}$ away from resonance, so that resonant absorption is
avoided at the natural resonance $\Delta_{0}=0$.

The feedback associated with these two stable solutions are weakly negative
and hence, according to Eq.~(\ref{FLUCTUATION2}), do not appreciably change
the nuclear field fluctuation. For example, at $\Delta_{0}=10$ $\mathrm{ns}%
^{-1}$ [marked by empty and filled circles in Fig.~\ref{G_FEEDBACK_POSITIVE}%
(a)-\ref{G_FEEDBACK_POSITIVE}(c)], the widths of the two peaks of
$p^{(\mathrm{ss})}(s)$ [solid curve in \ref{G_FEEDBACK_POSITIVE}(d)] are not
appreciably changed relative to the peak of the thermal equilibrium
distribution [dotted curve in Fig.~\ref{G_FEEDBACK_POSITIVE}(d)].

\subsubsection{Weak positive feedback}

\label{SEC_EXAMPLE_WEAKPOSITIVE_FEEDBACK}

\begin{figure}[ptb]
\includegraphics[width=\columnwidth]{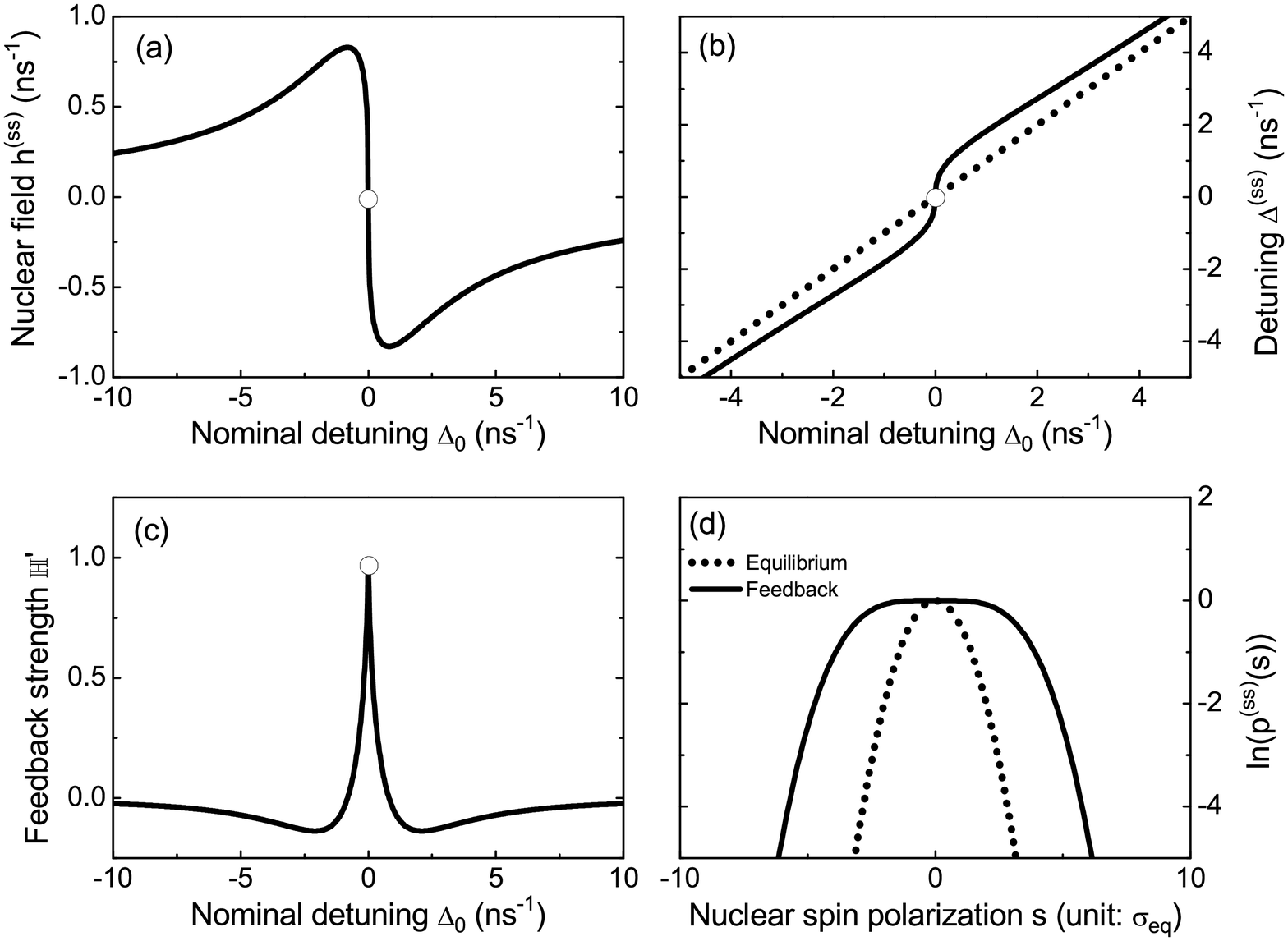}\caption{(color
online). The same as Fig.~\ref{G_FEEDBACK_POSITIVE}, except that $I=1/2$ and
$\gamma_{2}=1.8$ $\mathrm{ns}^{-1}$. In panel (b), the nominal detuning
(dotted straight line) is also shown for comparison. In panel (d), the peak of
$p^{(\mathrm{ss})}(s)$ correspond to the stable solution at $\Delta_{0}=0$ in
panels (a)-(c) (marked by empty circles).}%
\label{G_FEEDBACK_WEAKPOSITIVE}%
\end{figure}

For $\omega_{N}=0.1$ $\mathrm{ns}^{-1}$, to realize weak positive feedback
$\mathbb{H}^{\prime}(\Delta)\lesssim1$ near the resonance $\Delta=0$, we
decrease the magnitude of the feedback strength by considering nuclear
spin-1/2's instead of nuclear spin-9/2's and a larger hole dephasing rate
$\gamma_{2}=1.8$ $\mathrm{ns}^{-1}$.

In this case, since the feedback strength is small, the steady-state nuclear
field $h^{(\mathrm{ss})}$ [Fig.~\ref{G_FEEDBACK_WEAKPOSITIVE}(a)] is unique
and does not exhibit bistability. At $\Delta=0$ [marked by the empty circle in
Fig.~\ref{G_FEEDBACK_WEAKPOSITIVE}(a)-\ref{G_FEEDBACK_WEAKPOSITIVE}(c)], the
stable steady-state nuclear field $h^{(\mathrm{ss})}$ vanishes, so the
resonance condition $\Delta^{(\mathrm{ss})}=\Delta_{0}-h^{(\mathrm{ss})}$ and
hence resonant absorption is achieved at natural resonance $\Delta_{0}=0$.
However, due to the feedback strength $\mathbb{H}^{\prime}(0)\approx1$, a
slight change of $\Delta_{0}$ (or equivalently the pump frequency) away from
zero will drastically change the nuclear field
[Fig.~\ref{G_FEEDBACK_WEAKPOSITIVE}(a)] and hence the detuning $\Delta
^{(\mathrm{ss})}$ [Fig.~\ref{G_FEEDBACK_WEAKPOSITIVE}(b)] away from zero,
corresponding to large push-away from the natural resonance upon a slight
change of the pump frequency (Sec.~\ref{SEC_MEANFIELD_FEEDBACK}).

At $\Delta_{0}=0$, the feedback strength $\mathbb{H}^{\prime}(0)\approx1$
[Fig.~\ref{G_FEEDBACK_WEAKPOSITIVE}(c)]. According to Eq.~(\ref{FLUCTUATION2}%
), the nuclear field fluctuation is strongly enhanced, as can be seen from the
much wider peak of $p^{(\mathrm{ss})}(s)$ [solid curve in
Fig.~\ref{G_FEEDBACK_WEAKPOSITIVE}(d)] compared with the peak of the thermal
equilibrium distribution [dotted curve in Fig.~\ref{G_FEEDBACK_WEAKPOSITIVE}(d)].

\section{Conclusion}

We have developed a microscopic theory for the control of the nuclear field
dynamics by a general feedback loop mediated by the electron and/or the hole
(referred to as e-h system for brevity) under continuous wave pumping in a
quantum dot. This feedback loop consists of two steps. First, the nuclear
spins produce a quantum magnetic field $\hat{h}_{z}$ acting on the e-h system
[straight arrow in Fig.~\ref{G_SingleSpinJump}(a)] and establishes a $\hat
{h}_{z}$-dependent steady e-h state and hence $\hat{h}_{z}$-dependent e-h
fluctuation. Second, through a nonequilibrium fluctuation-dissipation
relation, the e-h fluctuation induces an irreversible nuclear spin population
flow [wavy arrow in Fig.~\ref{G_SingleSpinJump}(a)], which in turn changes the
nuclear field $\hat{h}_{z}$. By coupling the dynamics of individual nuclear
spins to the collective nuclear field $\hat{h}_{z}$, this feedback loop gains
control over the average nuclear field and the nuclear field fluctuation. This
control leads to three experimentally observed effects: (i) hysteresis in the
pump absorption strength; (ii)\ locking (avoidance) of the pump absorption
strength to (from) a certain value; (iii) suppression or amplification of the
nuclear field fluctuation, leading to prolonged or shortened electron spin
coherence time. By adiabatically eliminating the fast e-h motion in favor of
the slow nuclear field dynamics through the adiabatic approximation, we have
found that all these three effects can be quantified concisely by a single
nonlinear nuclear field feedback function $\mathbb{H}(h)$, which encapsulate
the mutual response between the e-h system and the nuclear field. A negative
(positive)\ feedback leads to locking (avoidance) of the pump absorption
strength and suppresses (amplifies) the nuclear field fluctuation. This
general theory is exemplified by considering a electron-hole-nuclei feedback
loop [Fig.~\ref{G_ENLOOP_HNLOOP}(b)] consisting of the hole-induced nuclear
spin flip through the non-collinear dipolar hyperfine interaction and the back
action of the nuclear field on the electron.

In the present work, we focus on the dynamics of the nuclear field on the time
scale of the nuclear spin relaxation, the longest time scale of the problem.
On a shorter time scale (much shorter than both the nuclear spin dephasing
time and the nuclear spin relaxation time, but still much longer than the time
scale of the e-h dynamics), we expect that a generalization of the adiabatic
approximation as used here could single out the dynamics of both the nuclear
spin coherence and the nuclear field, so that coherent nuclear spin dynamics
(e.g., nuclear spin coherent rotation and squeezing \cite{Rudner2011}) can be studied.

One limitation of the present treatment is that although the back action of
the diagonal coupling between the e-h system and the nuclear spins is treated
non-perturbatively, the off-diagonal coupling is treated by second-order
perturbation theory. This amounts to completely neglecting the back action of
the off-diagonal coupling on the e-h dynamics, e.g., the electron spin
relaxation due to the dynamic nuclear spin fluctuation through the
off-diagonal part of the electron-nuclear contact hyperfine
interaction.\cite{ChristPRB07,PetrovPRB09} This effect may be important when
the relaxation of the e-h system is dominated by the nuclear spins.

\begin{acknowledgments}
This research was supported by NSF (PHY 0804114, 1104446), ARO contract
W911NF-08-A0487, and U. S. Army Research Office MURI award W911NF0910406. We
thank D. G. Steel, X. Xu, B. Sun, R. B. Liu, W. Yao, A. H\"{o}gele, and A.
Imamoglu for helpful discussions. W. Y. thanks M. C. Zhang and Y. Wang for
helpful discussions.
\end{acknowledgments}

\appendix

\section{Nuclear spin dephasing by electron-hole fluctuation}

\label{APPENDIX_DEPHASING}

To show the nuclear spin dephasing induced by the e-h fluctuation through the
diagonal coupling $\hat{F}_{z}\hat{h}_{z}$, we divide the unperturbed
Hamiltonian $\hat{H}_{0}(t)$ into the uncoupled part $\hat{H}_{N}+\hat{H}%
_{eh}(t)$ and the coupling $\hat{F}_{z}\hat{h}_{z}$ and treat the latter by
perturbation theory. It is understood that the mean-field part $\hat{h}%
_{z}\operatorname*{Tr}_{eh}\hat{F}_{z}\hat{\rho}_{eh}^{(\mathrm{ss})}(t)$ of
the coupling $\hat{F}_{z}\hat{h}_{z}$ has been absorbed into the uncoupled
Hamiltonian. We start from the steady state $\hat{\rho}(t)=\hat{\rho}%
_{eh}^{(\mathrm{ss})}(t)\hat{\rho}_{N}^{(\mathrm{ss})}$ of the uncoupled
evolution
\[
\hat{u}(t,t_{0})=e^{-i\hat{H}_{N}(t-t_{0})}\mathcal{T}e^{-i\int_{t_{0}}%
^{t}\hat{H}_{eh}(t^{\prime})dt^{\prime}},
\]
define the interaction picture $\hat{O}^{\mathrm{I}}(t)\equiv\hat{u}^{\dagger
}(t,0)\hat{O}\hat{u}(t,0)$ and $\hat{\rho}^{\mathrm{I}}(t)\equiv\hat
{u}^{\dagger}(t,0)\hat{\rho}(t)\hat{u}(t,0)$, and turn on the coupling at
$t=0$. Treating the e-h system under continuous pumping as a nonequilibrium
bath and using the Born-Markov approximation\cite{GardinerQuantumNoise04} lead
to the equation for the nuclear spin density matrix $\hat{\rho}_{N}%
^{\mathrm{I}}(t)$ in the interaction picture:%
\[
\frac{d}{dt}\hat{\rho}_{N}^{\mathrm{I}}(t)=-\int_{0}^{t}dt^{\prime}[\hat
{F}_{z}^{\mathrm{I}}(t)\hat{h}_{z},[\hat{F}_{z}^{\mathrm{I}}(t^{\prime}%
)\hat{h}_{z},\hat{\rho}_{eh}^{(\mathrm{ss})}(0)\hat{\rho}_{N}^{\mathrm{I}%
}(t)]].
\]
By neglecting the imaginary part (corresponding to second-order nuclear spin
energy shift induced by the diagonal coupling) of the e-h fluctuation function
$\operatorname*{Tr}\nolimits_{eh}\hat{\rho}_{eh}^{(\mathrm{ss})}(0)\hat{F}%
_{z}^{\mathrm{I}}(t)\hat{F}_{z}^{\mathrm{I}}(t^{\prime})$, the above equation
can be put into the Lindblad form for pure dephasing:%
\begin{equation}
\frac{d}{dt}\hat{\rho}_{N}^{\mathrm{I}}(t)=-g_{z,z}(t)\left(  \frac{\hat
{h}_{z}\hat{h}_{z}\hat{\rho}_{N}^{\mathrm{I}}(t)+\hat{\rho}_{N}^{\mathrm{I}%
}(t)\hat{h}_{z}\hat{h}_{z}}{2}-\hat{h}_{z}\hat{\rho}_{N}^{\mathrm{I}}%
(t)\hat{h}_{z}\right)  . \label{DEPHASING}%
\end{equation}
Here the pure dephasing rate%
\[
g_{z,z}(t)\equiv\int_{0}^{t}dt^{\prime}\ \operatorname*{Tr}\nolimits_{eh}%
\hat{\rho}_{eh}^{(\mathrm{ss})}(0)\{\hat{F}_{z}^{\mathrm{I}}(t),\hat{F}%
_{z}^{\mathrm{I}}(t^{\prime})\}\sim||F_{z}||^{2}T_{eh},
\]
is determined by the e-h fluctuation, with $T_{eh}$ being its characteristic
decay time. In the product basis $|\mathbf{m}\rangle\equiv\otimes_{j}%
|m_{j}\rangle_{j}$ of the eigenstates $|m_{j}\rangle_{j}$ of individual
nuclear spin, Eq.~(\ref{DEPHASING}) becomes%
\[
\frac{d}{dt}\rho_{\mathbf{m},\mathbf{n}}^{\mathrm{I}}(t)=-\frac{1}{2}%
g_{z,z}(t)(\langle\mathbf{m}|\hat{h}_{z}|\mathbf{m}\rangle-\langle
\mathbf{n}|\hat{h}_{z}|\mathbf{n}\rangle)^{2}\rho_{\mathbf{m},\mathbf{n}%
}^{\mathrm{I}}(t),
\]
which shows that the coherence between two nuclear spin states $|\mathbf{m}%
\rangle$ and $|\mathbf{n}\rangle$ decays with rate $\sim(\langle
\mathbf{m}|\hat{h}_{z}|\mathbf{m}\rangle-\langle\mathbf{n}|\hat{h}%
_{z}|\mathbf{n}\rangle)^{2}T_{eh}$. For example, the coherence $\langle\hat
{I}_{j}^{\pm}\rangle$ of the $j$th nuclear spin decays with rate $\sim
a_{j,z}^{2}T_{eh}$. The inter-spin coherences $\langle\hat{I}_{i}^{\pm}\hat
{I}_{j}^{\pm}\rangle$ and $\langle\hat{I}_{i}^{\pm}\hat{I}_{j}^{\mp}\rangle$
decay with rates $\sim(a_{i,z}+a_{j,z})^{2}T_{eh}$ and $\sim(a_{i,z}%
-a_{j,z})^{2}T_{eh}$, respectively. Note that the inter-spin coherence
$\langle\hat{I}_{i}^{\pm}\hat{I}_{j}^{\mp}\rangle$ does not decay if
$a_{i,z}=a_{j,z}$. For the nuclear spin dynamics driven by an electron through
the contact hyperfine interaction [Eq.~(\ref{CONTACTHYPERFINE})], we have
$a_{j,z}=a_{j,e}\sim1-10\ \mathrm{\mu s}^{-1}$ for a quantum dot with
$N=10^{4}-10^{6}$ nuclear spins (see Table I). Using $T_{eh}\sim1/\gamma
_{1,2}\sim1\ \mathrm{ns}$, the single nuclear spin dephasing rate $\sim
a_{j,e}^{2}T_{eh}\sim0.01-1$ $\mathrm{ms}$. The persistent inter-spin
coherences for $a_{i,e}=a_{j,e}$ are manifested as nuclear spin dark states,
which limit the efficiency of dynamic nuclear
polarization.\cite{ImamogluPRL03,ChristPRB07,PetrovPRB09} Here to focus on the
nuclear spin feedback effect, we assume that persistent inter-spin coherences
have been removed, e.g., through e-h wave function
modulation.\cite{ChristPRB07,PetrovPRB09}

\section{Decoupling nuclear field dynamics from e-h dynamics through adiabatic
approximation}

\label{APPENDIX_RELAXATION}

The essential assumption of the adiabatic approximation is that on the time
scale $T_{1,N}$ of the nuclear spin flip driven by the off-diagonal coupling
$\hat{V}_{\mathrm{nd}}(t)$ [Eq.~(\ref{VND})], a classically correlated steady
state $\hat{\rho}_{eh}^{(\mathrm{ss})}(\hat{h}_{z},t)\hat{P}(t)$ is
instantaneously established by the unperturbed evolution driven by $\hat
{H}_{0}(t)$ [Eq.~(\ref{H0})]. To incorporate the influence of the nuclear spin
flip on the evolution of the diagonal part $\hat{P}(t)$ of the nuclear spin
density matrix, we start from the classically correlated state $\hat{\rho
}_{eh}^{(\mathrm{ss})}(\hat{h}_{z},0)\hat{P}(0)$ and turn on the off-diagonal
coupling at $t=0$. The density matrix $\hat{\rho}^{\mathrm{I}}(t)\equiv\hat
{U}_{0}^{\dagger}(t,0)\hat{\rho}(t)\hat{U}_{0}(t,0)$ in the interaction
picture obeys%
\begin{equation}
\frac{d}{dt}\hat{\rho}^{\mathrm{I}}(t)=-i[\hat{V}_{\mathrm{nd}}^{\mathrm{I}%
}(t),\hat{\rho}^{\mathrm{I}}(t)], \label{EOM_INTP}%
\end{equation}
where%
\[
\hat{U}_{0}(t,t_{0})\equiv\mathcal{T}e^{-i\int_{t_{0}}^{t}\hat{H}%
_{0}(t^{\prime})dt^{\prime}}%
\]
is the unperturbed evolution and
\[
\hat{V}_{\mathrm{nd}}^{\mathrm{I}}(t)=\hat{U}_{0}^{\dagger}(t,0)\hat
{V}_{\mathrm{nd}}(t)\hat{U}_{0}(t,0)\equiv\hat{F}_{+}^{\mathrm{I}}(t)\hat
{h}_{+}^{\mathrm{I}}(t)+\hat{F}_{-}^{\mathrm{I}}(t)\hat{h}_{-}^{\mathrm{I}%
}(t)
\]
is the operator in the interaction picture. For $|\omega_{N}|\gg|a_{j,\alpha
}|$, we can neglect the shift of nuclear Zeeman frequency induced by the
diagonal coupling, so that $\hat{h}_{+}^{\mathrm{I}}(t)\approx\sum_{j}%
a_{j,+}\hat{I}_{j,+}e^{i\omega_{j,N}t}$. Iterating Eq.~(\ref{EOM_INTP}) once
and transforming back to the Schr\"{o}dinger picture yields the exact equation%
\begin{align}
\frac{d}{dt}\hat{\rho}(t)  &  =-i[\hat{H}_{0}(t),\hat{\rho}(t)]-i[\hat
{V}_{\mathrm{nd}}(t),\hat{\rho}_{eh}^{(\mathrm{ss})}(\hat{h}_{z},t)\hat
{P}(0)]\label{EOM1}\\
&  -\int_{0}^{t}dt^{\prime}[\hat{V}_{\mathrm{nd}},\hat{U}_{0}(t,t^{\prime
})[\hat{V}_{\mathrm{nd}},\hat{\rho}(t^{\prime})]\hat{U}_{0}^{\dagger
}(t,t^{\prime})].\nonumber
\end{align}
Then, using the adiabatic approximation, we replace $\hat{\rho}(t^{\prime})$
by $\hat{\rho}_{eh}^{(\mathrm{ss})}(\hat{h}_{z},t^{\prime})\hat{P}(t^{\prime
})$ on the right hand side. Since the decay of the e-h fluctuation functions
\[
C_{a,b}(\hat{h}_{z},t,t^{\prime})\equiv\operatorname*{Tr}\nolimits_{eh}\hat
{F}_{a}^{\mathrm{I}}(t)\hat{F}_{b}^{\mathrm{I}}(t^{\prime})\hat{\rho}%
_{eh}^{(\mathrm{ss})}(\hat{h}_{z},0)
\]
is much faster than the motion of $\hat{P}(t^{\prime})$, we further replace
$\hat{\rho}_{eh}^{(\mathrm{ss})}(\hat{h}_{z},t^{\prime})\hat{P}(t^{\prime})$
by $\hat{\rho}_{eh}^{(\mathrm{ss})}(\hat{h}_{z},t^{\prime})\hat{P}(t)$ and
obtain
\begin{align*}
\frac{d}{dt}\hat{\rho}(t)  &  =-i[\hat{H}_{0}(t),\hat{\rho}(t)]-i[\hat
{V}_{\mathrm{nd}}(t),\hat{\rho}_{eh}^{(\mathrm{ss})}(\hat{h}_{z},t)\hat
{P}(0)]\\
&  -\int_{0}^{t}dt^{\prime}[\hat{V}_{\mathrm{nd}},[\hat{U}_{0}(t,t^{\prime
})\hat{V}_{\mathrm{nd}}\hat{U}_{0}^{\dagger}(t,t^{\prime}),\hat{\rho}%
_{eh}^{(\mathrm{ss})}(\hat{h}_{z},t)\hat{P}(t)]].
\end{align*}
Finally, tracing over the e-h degrees of freedom and taking the diagonal part
of the nuclear spin density matrix yields an equation of motion for $\hat
{P}(t)$:%
\begin{align}
\frac{d}{dt}\hat{P}(t)  &  =-\sum_{j}a_{j,+}^{2}\int_{0}^{t}dt^{\prime
}\{e^{-i\omega_{j,N}(t-t^{\prime})}[\hat{I}_{j,-},\hat{I}_{j,+}C_{+,-}(\hat
{h}_{z},t,t^{\prime})\hat{P}(t)]\nonumber\\
&  +e^{i\omega_{j,N}(t-t^{\prime})}[\hat{I}_{j,+},\hat{I}_{j,-}C_{-,+}(\hat
{h}_{z},t,t^{\prime})\hat{P}(t)]+h.c.\}. \label{EOM2}%
\end{align}
The above equation shows that the nuclear spin dynamics is driven by the e-h
fluctuation, which in turn is controlled by the continuous pumping and the
nuclear field $\hat{h}_{z}$.

If the time dependence of the effective e-h Hamiltonian $\hat{H}_{eh}(\hat
{h}_{z},t)\equiv\hat{H}_{eh}(t)+\hat{F}_{z}\hat{h}_{z}$ can be eliminated by a
rotating wave transformation
\begin{subequations}
\label{ROT}%
\begin{align}
\hat{\rho}_{eh}^{\mathrm{rot}}(t)  &  =\hat{R}(t)\hat{\rho}_{eh}(t)\hat
{R}^{-1}(t),\label{ROTA}\\
\hat{H}_{eh}^{\mathrm{rot}}(\hat{h}_{z})  &  =\hat{R}(t)\hat{H}_{eh}(\hat
{h}_{z},t)\hat{R}^{-1}(t)+i\frac{d\hat{R}(t)}{dt}\hat{R}^{-1}(t), \label{ROTB}%
\end{align}
then $\hat{\rho}_{eh}^{(\mathrm{ss})}(\hat{h}_{z},0)$ is the time-independent
steady state in the rotating frame, i.e., $\hat{\rho}_{eh}^{(\mathrm{ss}%
)}(\hat{h}_{z},0)$ commutes with $\hat{H}_{eh}^{\mathrm{rot}}(\hat{h}_{z})$,
so that
\end{subequations}
\begin{align*}
\hat{U}_{0}(t,t_{0})  &  =e^{-i\hat{H}_{N}(t-t_{0})}\hat{R}^{-1}%
(t)e^{-i\hat{H}_{eh}^{\mathrm{rot}}(\hat{h}_{z})(t-t_{0})}\hat{R}(t_{0}),\\
C_{a,b}(\hat{h}_{z},t,t^{\prime})  &  =\operatorname*{Tr}\nolimits_{eh}%
e^{i\hat{H}_{eh}^{\mathrm{rot}}(\hat{h}_{z})(t-t^{\prime})}\hat{F}%
_{a}^{\mathrm{rot}}(t)e^{-i\hat{H}_{eh}^{\mathrm{rot}}(\hat{h}_{z}%
)(t-t^{\prime})}\\
&  \times\hat{F}_{b}^{\mathrm{rot}}(t^{\prime})\hat{\rho}_{eh}^{(\mathrm{ss}%
)}(\hat{h}_{z},0),
\end{align*}
where $\hat{O}^{\mathrm{rot}}(t)\equiv\hat{R}(t)\hat{O}(t)\hat{R}^{-1}(t)$ is
the operator in the rotating frame. In the simplest case, the e-h operator
$\hat{F}_{-}(t)=\hat{F}_{+}^{\dagger}(t)$ has a definite frequency $\hat
{F}_{-}^{\mathrm{rot}}(t)=e^{-i\omega_{0}t}\hat{F}_{-}$ in the rotating frame,
then $C_{\mp,\pm}(\hat{h}_{z},t,t^{\prime})$ are invariant under simultaneous
temporal translation:\ $t\rightarrow t+\tau$ and $t^{\prime}\rightarrow
t^{\prime}+\tau$. Generally, $\hat{F}_{-}^{\mathrm{rot}}(t)$ can be expanded
into different frequency components%
\[
\hat{F}_{-}^{\mathrm{rot}}(t)=\sum_{\alpha}\hat{F}_{-,\alpha}e^{-i\omega
_{\alpha}t}%
\]
and the e-h fluctuation functions
\[
C_{\mp,\pm}(\hat{h}_{z},t,t^{\prime})=\sum_{\alpha\beta}e^{-i(\omega_{\alpha
}-\omega_{\beta})t}C_{\mp\alpha,\pm\beta}(\hat{h}_{z},t-t^{\prime})
\]
contain the interference terms%
\begin{align*}
&  C_{\mp\alpha,\pm\beta}(\hat{h}_{z},\tau)\\
&  =e^{-i\omega_{\beta}\tau}\operatorname*{Tr}\nolimits_{eh}e^{i\hat{H}%
_{eh}^{\mathrm{rot}}(\hat{h}_{z})\tau}\hat{F}_{\mp,\alpha}e^{-i\hat{H}%
_{eh}^{\mathrm{rot}}(\hat{h}_{z})\tau}\hat{F}_{\pm,\beta}^{(\mathrm{ss})}%
\hat{\rho}_{eh}^{(\mathrm{ss})}(\hat{h}_{z},0).
\end{align*}
However, on the time scale $T_{1,N}$ of nuclear spin relaxation, the rapidly
oscillating phase factor $e^{-i(\omega_{\alpha}-\omega_{\beta})t}$ averages
out the interference terms if $|\omega_{\alpha}-\omega_{\beta}|T_{1,N}\gg1$,
which is satisfied for $\alpha\neq\beta$ under typical experimental
conditions. Neglecting the interference terms restores the temporal
translational invariance of the e-h fluctuation functions $\hat{C}_{\mp,\pm
}(\hat{h}_{z},\tau)=$ $\sum_{\alpha}\hat{C}_{\mp\alpha,\pm\alpha}(\hat{h}%
_{z},\tau)$. Note that even when the time dependence of $\hat{H}_{eh}(\hat
{h}_{z},t)$ cannot be eliminated by a rotating wave transformation, a similar
reasoning can be used to show that the steady-state fluctuation functions
$\hat{C}_{\mp,\pm}(\hat{h}_{z},t,t^{\prime})$ (with $t,t^{\prime}%
\rightarrow\infty$) is invariant under temporal translation if $|\omega
_{\alpha}-\omega_{\beta}|T_{1,N}\gg1$ is satisfied for two arbitrary
characteristic frequencies of $\hat{F}_{+}^{\mathrm{I}}(t)$.

For the e-h fluctuation functions being invariant under temporal translations,
by further neglecting the second-order energy correction of the nuclear spins
induced by the off-diagonal coupling, Eq.~(\ref{EOM2}) simplifies to
Eq.~(\ref{RATEEQ}).

\section{Derivation of Fokker-Planck equation}

\label{APPENDIX_FP}

The equation of motion of $p(s,t)$ follows from Eq.~(\ref{RATEEQ}) as%
\begin{align*}
\frac{\partial}{\partial t}p(s,t)  &  =-NI[W_{+}(h_{\max}s)\operatorname*{Tr}%
\delta_{\hat{s},s}(\hat{K}-s)\hat{P}(t)\\
&  -W_{+}(h_{\max}(s-a))\operatorname*{Tr}(\hat{K}-(s-a))\delta_{\hat{s}%
,s-a}\hat{P}(t)]\\
&  -NI[W_{-}(h_{\max}s)\operatorname*{Tr}\delta_{\hat{s},s}(\hat{K}+s)\hat
{P}(t)\\
&  -W_{-}(h_{\max}(s+a))\operatorname*{Tr}(\hat{K}+s+a)\delta_{\hat{s}%
,s+a}\hat{P}(t)],
\end{align*}
where $N$ is the number of nuclear spins in the QD, $a\equiv1/(NI)$ is the
change of $\hat{s}$ by each nuclear spin flip, and%
\[
\hat{K}\equiv\frac{1}{NI}\sum_{j}(\hat{I}_{j,x}^{2}+\hat{I}_{j,y}^{2}).
\]
For nuclear spin-1/2's or weak nuclear spin polarization $|s_{0}(h)|\ll1$, the
transverse fluctuation of each individual nuclear spin is not influenced by
its longitudinal polarization, then $\hat{K}$ can be replaced with $2(I+1)/3$
and we obtain a closed equation for $p(s,t)$:%
\begin{align}
\frac{\partial}{\partial t}p(s,t)  &  =-[G_{+}(s)p(s,t)-G_{+}%
(s-a)p(s-a,t)]\label{EOM_PST}\\
&  -[G_{-}(s)p(s,t)-G_{-}(s+a)p(s+a,t)],\nonumber
\end{align}
where%
\[
G_{\pm}(s)\equiv NIW_{\pm}(h_{\max}s)\left(  \frac{2(I+1)}{3}\mp s\right)  .
\]
For $N\gg1$, we expand Eq.~(\ref{EOM_PST}) up to the second order of the small
quantity $a$ and obtain the Fokker-Planck equation Eq.~(\ref{FOKKER_PLANCK}).

For the general situation, the equation of motion for $p(s,t)$ is not closed.
In this case, to quantify the fluctuation of nuclear spin-$I$'s, we define the
population number operator%
\[
\hat{N}_{m}\equiv\sum_{j=1}^{N}\left\vert m\right\rangle _{j}\left\langle
m\right\vert
\]
to count the number of nuclear spins in the $m$th single spin eigenstate
$\left\vert m\right\rangle $ for $m=-I,-(I-1),\cdots,I$ and the population
$\hat{\mathbf{N}}\equiv\lbrack\hat{N}_{-I},\cdots,\hat{N}_{I}]^{T}$ to
characterizes the state of the nuclear spins. The information about the
nuclear spin fluctuation is contained in the population number distribution
\[
p(\mathbf{N},t)\equiv\operatorname*{Tr}\hat{P}(t)\prod_{m}\delta_{\hat{N}%
_{m},N_{m}}\equiv\operatorname*{Tr}\hat{P}(t)\delta_{\hat{\mathbf{N}%
},\mathbf{N}},
\]
where $\mathbf{N}\equiv\lbrack N_{-I},\cdots,N_{I}]^{T}$. Straightforward
algebra shows that $p(\mathbf{N},t)$ obeys the equation of motion%
\begin{align}
\frac{\partial}{\partial t}p(\mathbf{N},t)  &  =-\sum_{m}N_{m}\eta_{m}%
^{2}W_{+}(h(\mathbf{N}))p(\mathbf{N},t)\nonumber\\
&  +\sum_{m}(N_{m}+1)\eta_{m}^{2}W_{+}(h(\mathbf{N}^{(m,m+1)}))p(\mathbf{N}%
^{(m,m+1)},t)\nonumber\\
&  -\sum_{m}N_{m+1}\eta_{m}^{2}W_{-}(h(\mathbf{N}))p(\mathbf{N},t)\nonumber\\
&  +\sum_{m}(N_{m+1}+1)\eta_{m}^{2}W_{-}(h(\mathbf{N}^{(m+1,m)}))p(\mathbf{N}%
^{(m+1,m)},t), \label{EOM_PNT}%
\end{align}
where $\eta_{m}\equiv\left\langle m+1\left\vert I^{+}\right\vert
m\right\rangle $,
\[
h(\mathbf{N})\equiv a_{z}\sum_{m}mN_{m}%
\]
is the nuclear field produced by nuclear spins in a state characterized by
population numbers $\mathbf{N}$,
\[
\mathbf{N}^{(m,m+1)}\equiv\lbrack\cdots,N_{m}+1,N_{m+1}-1,\cdots]^{T}%
\]
is obtained from $\mathbf{N}$ by flipping one nuclear spin from state
$\left\vert m+1\right\rangle $ to state $\left\vert m\right\rangle $, and
\[
\mathbf{N}^{(m+1,m)}\equiv\lbrack\cdots,N_{m}-1,N_{m+1}+1,\cdots]^{T}%
\]
is obtained from $\mathbf{N}$ by flipping one nuclear spin from state
$\left\vert m\right\rangle $ to state $\left\vert m+1\right\rangle $. The
corresponding Overhauser shift $h(\mathbf{N}^{(m,m+1)})=h(\mathbf{N})-a_{z}$
and $h(\mathbf{N}^{(m+1,m)})=h(\mathbf{N})+a_{z}$. Equation (\ref{EOM_PNT})
describes the population flow of the nuclear spins induced by the e-h
fluctuation. The first two terms of Eq.~(\ref{EOM_PNT}) come from the jump of
one nuclear spin from $\left\vert m\right\rangle $ to $\left\vert
m+1\right\rangle $, which changes the nuclear spin population from
$\mathbf{N}$ to $\mathbf{N}^{(m+1,m)}$ (the first term) or from $\mathbf{N}%
^{(m,m+1)}$ to $\mathbf{N}$ (the second term). The last two terms come from
the jump of one nuclear spin from $\left\vert m+1\right\rangle $ to
$\left\vert m\right\rangle $, which changes the nuclear spin population from
$\mathbf{N}$ to $\mathbf{N}^{(m,m+1)}$ (the third term) or from $\mathbf{N}%
^{(m+1,m)}$ to $\mathbf{N}$ (the fourth term). For $N\gg1$, a second-order
Taylor expansion can be used to transform Eq.~(\ref{EOM_PNT}) into a
multi-variable Fokker-Planck equation.

As an example, for $I=1$, we define $x=N_{+1}/N$ and $y=N_{-1}/N$, where
$N=N_{-1}+N_{0}+N_{+1}$ is the total number of nuclear spins in the quantum
dot. The equation of motion for the distribution function $q(x,y,t)\equiv
p(N_{-1},N_{0},N_{+1})$ follows from Eq.~(\ref{EOM_PNT}) as
\begin{align*}
\frac{\partial}{\partial t}q(x,y,t)  &  =-g_{y+}(x,y)p(x,y,t)+g_{y+}%
(x,y+a)p(x,y+a,t)\\
&  \quad-g_{x-}(x,y)p(x,y,t)+g_{x-}(x-a,y)p(x-a,y,t)\\
&  \quad-g_{y-}(x,y)p(x,y,t)+g_{y-}(x,y-a)p(x,y-a,t)\\
&  \quad-g_{x+}(x,y)p(x,y,t)+g_{x+}(x+a,y)p(x+a,y,t),
\end{align*}
where%
\begin{align*}
g_{x+}(x,y)  &  =2NxW_{-}(h_{\max}(x-y)),\\
g_{x-}(x,y)  &  =2N(1-x-y)W_{+}(h_{\max}(x-y)),\\
g_{y+}(x,y)  &  =2NyW_{+}(h_{\max}(x-y)),\\
g_{y-}(x,y)  &  =2N(1-x-y)W_{-}(h_{\max}(x-y)).
\end{align*}
Through a second-order Taylor expansion, we obtain the Fokker-Planck equation%
\begin{align*}
\frac{\partial}{\partial t}q(x,y,t)  &  =-\frac{\partial}{\partial x}\left[
v_{x}(x,y)p(x,y,t)-\frac{\partial}{\partial x}D_{xx}(x,y)p(x,y,t)\right] \\
&  -\frac{\partial}{\partial y}\left[  v_{y}(x,y)p(x,y,t)-\frac{\partial
}{\partial y}D_{yy}(x,y)p(x,y,t)\right]
\end{align*}
where%
\begin{align*}
v_{x}(x,y)  &  \equiv a\left[  g_{x-}(x,y)-g_{x+}(x,y)\right]  ,\\
v_{y}(x,y)  &  \equiv a\left[  g_{y-}(x,y)-g_{y+}(x,y)\right]  ,\\
D_{xx}(x,y)  &  \equiv\frac{1}{2}a^{2}\left[  g_{x-}(x,y)+g_{x+}(x,y)\right]
,\\
D_{yy}(x,y)  &  \equiv\frac{1}{2}a^{2}\left[  g_{y+}(x,y)+g_{y-}(x,y)\right]
.
\end{align*}

\section{Evaluation of e-h fluctuation function $C(\Delta,\nu)$}

\label{APPENDIX_QR}

In the absence of the nuclear spins, the e-h fluctuation function becomes%

\begin{align*}
C(\Delta_{0},\nu)  &  \equiv\int_{-\infty}^{\infty}e^{i\nu t}%
dt\ \operatorname*{Tr}\nolimits_{eh}\hat{\sigma}_{1,1}^{\mathrm{I}}%
(t)\hat{\sigma}_{1,1}^{\mathrm{I}}(0)\hat{\rho}_{eh}^{(\mathrm{ss})}(0)\\
&  =\int_{0}^{\infty}e^{i\nu t}dt\ \operatorname*{Tr}\nolimits_{eh}\hat
{\sigma}_{1,1}^{\mathrm{I}}(t)\hat{\sigma}_{1,1}^{\mathrm{I}}(0)\hat{\rho
}_{eh}^{(\mathrm{ss})}(0)+h.c.,
\end{align*}
where $\hat{\sigma}_{1,1}^{\mathrm{I}}(t)$ is driven by the free e-h evolution
$\hat{U}_{eh}(t)$ [Eq.~(\ref{UEH})] and $\hat{\rho}_{eh}^{(\mathrm{ss})}(0)$
is the steady-state of the e-h system in the absence of the nuclear spins.
Below we evaluate $C(\Delta_{0},\nu)$, so that $C(\Delta,\nu)$ is obtained by
replacing $\Delta_{0}$ with $\Delta$.

First we define a three-component operator $\mathbf{\hat{X}}(t)\equiv
\lbrack\hat{\sigma}_{1,1}^{\mathrm{I}}(t),e^{i\omega t}\hat{\sigma}%
_{0,1}^{\mathrm{I}}(t),e^{-i\omega t}\hat{\sigma}_{1,0}^{\mathrm{I}}(t)]^{T}$
and its average
\[
\langle\mathbf{\hat{X}}(t)\mathbf{\rangle}_{(\mathrm{ss})}\equiv
\operatorname*{Tr}\nolimits_{eh}\mathbf{\hat{X}}(t)\hat{\rho}_{eh}%
^{(\mathrm{ss})}(0)
\]
over the steady e-h state $\hat{\rho}_{eh}^{(\mathrm{ss})}(0)$. Since the
coupling $\hat{H}_{\mathrm{damp}}$ to the vacuum electromagnetic fluctuation
induces the hole relaxation $\left\vert 1\right\rangle \rightarrow\left\vert
0\right\rangle $ with rate $\gamma_{1}$ and hole dephasing with total rate
$\gamma_{2}$ in the Lindblad form, the equation of motion of $\langle
\mathbf{\hat{X}}(t)\mathbf{\rangle}_{(\mathrm{ss})}$ is given by
\begin{align*}
\frac{d}{dt}\langle\mathbf{\hat{X}}(t)\mathbf{\rangle}_{(\mathrm{ss})}  &
=-\mathbf{A[}\langle\mathbf{\hat{X}}(t)\mathbf{\rangle}_{(\mathrm{ss}%
)}-\mathbf{A}^{-1}\mathbf{B}]\\
&  \equiv-\mathbf{A[}\langle\mathbf{\hat{X}}(t)\mathbf{\rangle}_{(\mathrm{ss}%
)}-\langle\mathbf{\hat{X}}(+\infty)\mathbf{\rangle}_{(\mathrm{ss})}]
\end{align*}
for $t>0$, where
\[
\mathbf{A}=%
\begin{bmatrix}
\gamma_{1} & -i\Omega_{R}/2 & i\Omega_{R}/2\\
-i\Omega_{R} & i(\Delta_{0}-i\gamma_{2}) & 0\\
i\Omega_{R} & 0 & -i(\Delta_{0}+i\gamma_{2})
\end{bmatrix}
,
\]
and $\mathbf{B}=[0,-i\Omega_{R}/2,i\Omega_{R}/2]^{T}$. According to the
quantum regression theorem,\cite{ScullyQuantumOptics} the fluctuation
functions $\langle\mathbf{\hat{X}}(t)\hat{\sigma}_{1,1}\mathbf{\rangle
}_{(\mathrm{ss})}$ obey a similar equation%
\[
\frac{d}{dt}\langle\mathbf{\hat{X}}(t)\hat{\sigma}_{1,1}\mathbf{\rangle
}_{(\mathrm{ss})}=-\mathbf{A[}\langle\mathbf{\hat{X}}(t)\hat{\sigma}%
_{1,1}\mathbf{\rangle}_{(\mathrm{ss})}-\langle\mathbf{\hat{X}}(+\infty
)\mathbf{\rangle}_{(\mathrm{ss})}\langle\hat{\sigma}_{1,1}\rangle
_{(\mathrm{ss})}\mathbf{],}%
\]
from which we obtain%
\[
\langle\mathbf{\hat{X}}(t)\hat{\sigma}_{1,1}\mathbf{\rangle}_{(\mathrm{ss}%
)}=\langle\mathbf{\hat{X}\rangle}_{(\mathrm{ss})}\langle\hat{\sigma}%
_{1,1}\rangle_{(\mathrm{ss})}+e^{-\mathbf{A}t}\left[  \langle\mathbf{\hat{X}%
}\hat{\sigma}_{1,1}\mathbf{\rangle}_{(\mathrm{ss})}-\langle\mathbf{\hat
{X}\rangle}_{(\mathrm{ss})}\langle\hat{\sigma}_{1,1}\rangle_{(\mathrm{ss}%
)}\right]  ,
\]
where we have used $\langle\mathbf{\hat{X}}(+\infty)\mathbf{\rangle
}_{(\mathrm{ss})}=\langle\mathbf{\hat{X}\rangle}_{(\mathrm{ss})}$ for the
steady-state average. As a result, $C(\Delta_{0},\nu)$ is given by the first
element of
\begin{align*}
&  \int_{0}^{\infty}e^{i\nu t}dt\ \mathbf{[}\langle\mathbf{\hat{X}}%
(t)\hat{\sigma}_{1,1}\mathbf{\rangle}_{(\mathrm{ss})}-\langle\mathbf{\hat
{X}\rangle}_{(\mathrm{ss})}\langle\hat{\sigma}_{1,1}\rangle_{(\mathrm{ss}%
)}\mathbf{]}+h.c.\\
&  =(\mathbf{A}-i\nu)^{-1}\mathbf{[}\langle\mathbf{\hat{X}}\hat{\sigma}%
_{1,1}\mathbf{\rangle}_{(\mathrm{ss})}-\langle\mathbf{\hat{X}\rangle
}_{(\mathrm{ss})}\langle\hat{\sigma}_{1,1}\rangle_{(\mathrm{ss})}%
\mathbf{]}+h.c.
\end{align*}

\bibliographystyle{apsrev4-1}
%\bibliography{e://dropbox/myLiterature}

\end{document}